# The Hierarchical Dynamical State of Molecular Gas from 3 to 300 pc in NGC 253

Elias K. Oakes[1], Christopher M. Faesi[1], Erik Rosolowsky[2], Adam K. Leroy[3,4], Simon C. O. Glover[5], Annie Hughes[6], Sharon E. Meidt[7], Eva Schinnerer[8], Jiayi Sun[9,25], Amirnezam Amiri[10], Ashley T. Barnes[11], Zein Bazzi[12], Ivana Bešlić[13], Frank Bigiel[12], Guillermo A. Blanc[14,15], Charlie Burton[2], Ryan Chown[3], Enrico Congiu[16], Daniel A. Dale[17], Simthembile Dlamini[18], Hao He[12], Eric W. Koch[19], Fu-Heng Liang[20], Jérôme Pety[13,21], Miguel Querejeta[22], Sumit K. Sarbadhicary[23], Sophia K. Stuber[8], Antonio Usero[22], and Thomas G. Williams[24]

[1] Department of Physics, University of Connecticut, 196A Auditorium Road, Storrs, CT 06269, USA; elias.oakes@uconn.edu
[2] Department of Physics, University of Alberta, Edmonton, Alberta, T6G 2E1, Canada
[3] Department of Astronomy, The Ohio State University, 140 West 18th Avenue, Columbus, OH 43210, USA
[4] Center for Cosmology and Astroparticle Physics (CCAPP), 191 West Woodruff Avenue, Columbus, OH 43210, USA
[5] Universität Heidelberg, Zentrum für Astronomie, Institut für Theoretische Astrophysik, Albert-Ueberle-Str 2, D-69120 Heidelberg, Germany
[6] IRAP/OMP/Université de Toulouse, 9 Av. du Colonel Roche, BP 44346, F-31028 Toulouse cedex 4, France
[7] Sterrenkundig Observatorium, Universiteit Gent, Krijgslaan 281 S9, B-9000 Gent, Belgium
[8] Max Planck Institut für Astronomie, Königstuhl 17, 69117 Heidelberg, Germany
[9] Department of Astrophysical Sciences, Princeton University, 4 Ivy Lane, Princeton, NJ 08544, USA
[10] Department of Physics, University of Arkansas, 226 Physics Building, 825 West Dickson Street, Fayetteville, AR 72701, USA
[11] European Southern Observatory, Karl-Schwarzschild-Strasse 2, 85748 Garching bei München, Germany
[12] Argelander-Institut für Astronomie, University of Bonn, Auf dem Hügel 71, 53121 Bonn, Germany
[13] LUX, Observatoire de Paris, PSL Research University, CNRS, Sorbonne Universités, 75014 Paris, France
[14] Observatories of the Carnegie Institution for Science, 813 Santa Barbara Street, Pasadena, CA 91101, USA
[15] Departamento de Astronomía, Universidad de Chile, Camino del Observatorio 1515, Las Condes, Santiago, Chile
[16] European Southern Observatory (ESO), Alonso de Córdova 3107, Casilla 19, Santiago 19001, Chile
[17] Department of Physics and Astronomy, University of Wyoming, Laramie, WY 82071, USA
[18] Department of Astronomy, University of Cape Town, Rondebosch 7701, South Africa
[19] Center for Astrophysics | Harvard & Smithsonian, 60 Garden Street, Cambridge, MA 02138, USA
[20] Astronomisches Rechen-Institut, Zentrum für Astronomie der Universität Heidelberg, Mönchhofstraße 12-14, 69120 Heidelberg, Germany
[21] IRAM, 300 rue de la Piscine, 38400 Saint Martin d'Héres, France
[22] Observatorio Astronómico Nacional (IGN), C/Alfonso XII 3, Madrid E-28014, Spain
[23] Department of Physics and Astronomy, Johns Hopkins University, Baltimore, MD 21218, USA
[24] Sub-department of Astrophysics, Department of Physics, University of Oxford, Keble Road, Oxford OX1 3RH, UK

## Abstract

Understanding how the dynamical state of the interstellar medium (ISM) changes across spatial scales can provide important insights into how the gas is organized and ultimately collapses to form stars. To this end, we present ALMA $^{12}$CO(2–1) observations at 7 pc ($0\overset{''}{.}4$) spatial resolution across a 1.4 kpc × 5.6 kpc ($1\overset{'}{.}3 \times 1\overset{'}{.}3$) region located in the disk of the nearby ($D = 3.5$ Mpc), massive, star-forming galaxy NGC 253. We decompose this emission with a hierarchical, multiscale dendrogram algorithm to identify 2463 structures with deconvolved sizes ranging from ∼3 to 300 pc, complete to a limiting mass of $10^4\,M_\odot$. By comparing the virial parameter of these structures against physical properties including size, mass, surface density, velocity dispersion, and hierarchical position, we carry out a comprehensive search for a preferred scale at which gravitationally bound structures emerge. Ultimately, we do not identify evidence of an emergent scale for bound objects in our data, nor do we find a significant correlation between the virial parameter and structure sizes. These findings suggest that simple observational estimates of gravitational binding cannot be used to define molecular clouds and emphasize the need for multiscale approaches to characterize the ISM.

*Materials only available in the online version of record: machine-readable table*

## 1. Introduction: Searching for Clouds in a Multiscale ISM

Star formation is ultimately driven by the collapse of gravitationally bound gas (R. B. Larson 2003; M. R. Krumholz & C. F. McKee 2005; C. F. McKee & E. C. Ostriker 2007). In the local Universe, this process occurs in dense, sub-parsec-scale molecular cores embedded within a hierarchy of progressively more diffuse interstellar medium (ISM) gas structures, notably molecular clouds and their "giant" counterparts (GMCs; see reviews by C. L. Dobbs & J. E. Pringle 2013; M. Heyer & T. M. Dame 2015; F. Motte et al. 2018; J. Ballesteros-Paredes et al. 2020; M. Chevance et al. 2020, 2023; E. Schinnerer & A. K. Leroy 2024; H. Beuther et al. 2025). Populations of cloud-scale (∼50–100 pc) objects have been extensively cataloged in the Milky Way (e.g., P. M. Solomon et al. 1987; T. S. Rice et al. 2016; M.-A. Miville-Deschênes et al. 2017) and across the galaxy population, with physical properties that couple closely to the large-scale galactic environment (e.g., A. K. Leroy et al. 2017; C. M. Faesi et al. 2018; S. E. Meidt et al. 2018; A. K. Leroy et al. 2021b; E. Rosolowsky et al. 2021; J. Sun et al. 2022). Theories such as global hierarchical collapse (E. Vázquez-Semadeni et al. 2019, 2024a) and inertial inflow (P. Padoan et al. 2020) posit mechanisms for concentrating this large-scale gas into star-forming cores at the centers of hub-filament systems (e.g., M. S. N. Kumar et al. 2020; D. Yang et al. 2023; S. Pan et al. 2024). To test these theories, however, it is essential to understand how the properties and dynamics of molecular gas structures vary *across* size scales. In doing so, we can establish where gravity begins to dominate and define at what scales bound structures may emerge.

[25] NASA Hubble Fellow.

To characterize the dynamical state of gas, a simple virial parameter, $\alpha_{\rm vir} \equiv 2K/U_{\rm G}$, has historically been used to approximate the balance between kinetic ($K$) and gravitational ($U_{\rm g}$) energy in the absence of other energy contributions (P. M. Solomon et al. 1987; F. Bertoldi & C. F. McKee 1992; C. F. McKee & E. G. Zweibel 1992). As gas with low $\alpha_{\rm vir}$ tends to be gravitationally bound, its value is closely tied to star formation in theoretical models (C. F. McKee & E. G. Zweibel 1992; M. R. Krumholz & C. F. McKee 2005; P. Padoan & Å. Nordlund 2011; C. Federrath & R. S. Klessen 2012; J. M. D. Kruijssen 2012; M. R. Krumholz et al. 2012; P. Hennebelle & G. Chabrier 2013; P. Padoan et al. 2017; J.-G. Kim et al. 2021, among others).

Historically, our view of the molecular ISM has considered the medium to be composed of distinct, often self-gravitating entities at characteristic scales of ∼50–100 pc: GMCs (N. Z. Scoville & P. M. Solomon 1975; P. M. Solomon et al. 1979; R. B. Larson 1981; P. M. Solomon et al. 1987; M. H. Heyer et al. 2001). While the concept of discrete GMCs is fairly ubiquitous in the field, it remains poorly defined in physical terms. Indeed, in more recent years, this view has been challenged by a greater focus on the ISM's dynamic and hierarchical nature, partly motivated by observations and simulations suggesting that clouds may be out-of-equilibrium structures (C. L. Dobbs et al. 2011; N. J. I. Evans et al. 2021; L. Liu et al. 2021; R. G. Treß et al. 2021) that fluctuate in their shape and mass and have relatively short lifetimes (e.g., B. G. Elmegreen 2000; L. Hartmann et al. 2001; J. Ballesteros-Paredes 2006; S. Dib et al. 2007; C. L. Dobbs & J. E. Pringle 2013; A. Hughes et al. 2013; S. E. Meidt et al. 2015; S. M. R. Jeffreson et al. 2021; M. Chevance et al. 2022, and discussions therein).

The GMC paradigm has framed much of the development and interpretation of structural decomposition algorithms applied to molecular gas position–position–velocity (PPV) data cubes. Various segmentation methods exist, from simple brightness thresholding (D. B. Sanders & I. F. Mirabel 1985; P. M. Solomon et al. 1987; T. M. Dame et al. 2001) to more sophisticated approaches that identify characteristic geometries (GAUSSCLUMPS, J. Stutzki & R. Guesten 1990), or associate neighboring voxels by their values (Clumpfind and CPROPS, J. P. Williams et al. 1994; E. Rosolowsky & A. Leroy 2006; E. Rosolowsky et al. 2021). However, these techniques often impose observationally driven scales on the resulting samples, among other methodological challenges (N. Schneider & K. Brooks 2004; A. Hughes et al. 2013; D. Colombo et al. 2015; A. K. Leroy et al. 2016). This motivates nonparametric approaches that make minimal assumptions about the organization and scales of molecular gas structures, either by avoiding cloud identification altogether (T. Sawada et al. 2012; A. K. Leroy et al. 2016; J. Sun et al. 2018; T. G. Williams et al. 2023) or by identifying structure across a multiscale hierarchy (E. W. Rosolowsky et al. 2008; A. Hughes et al. 2013; D. Colombo et al. 2015; G.-X. Li 2022; H. Yu & X. Hou 2022).

Given this complex landscape, whether molecular clouds are "real" remains an open question. Establishing the physical nature of GMCs would imply some characteristic size, mass, or density scale at which clouds distinguish themselves from their environments. Observationally, we might expect this decoupling to occur via their gravitational state ($\alpha_{\rm vir}$), turbulent statistics, or dynamical properties. Dendrograms, which identify structure on all possible scales (e.g., A. A. Goodman et al. 2009; R. Shetty et al. 2012; B. Burkhart et al. 2013; S. Storm et al. 2014; D. Colombo et al. 2015; R. Shen et al. 2024), make the search for an emergent physical GMC scale possible. Furthermore, studying the differential change in structure properties can uniquely circumvent the high uncertainties inherent to estimates of $\alpha_{\rm vir}$ (C. J. Lada et al. 2025; M. R. Krumholz et al. 2025). However, leveraging this approach requires observations with high physical and velocity resolution, wide areal coverage, and good sensitivity, key to accessing self-gravitating gas at small scales while placing these in the clear context of an extended molecular region. Due to line-of-sight blending and limited fields of view, such observations are challenging in the Milky Way and require targeting the nearest extragalactic systems.

In this work, we present unique ALMA observations of a region in NGC 253's disk that provides just this view. As the nearest massive star-forming galaxy observable by ALMA ($D = 3.5 \pm 0.2$ Mpc, S. Okamoto et al. 2024), NGC 253 provides the external perspective to the Milky Way of a massive barred spiral galaxy with a molecule-rich center (A. K. Leroy et al. 2015; N. Krieger et al. 2020). We targeted the particular $80'' \times 80''$ (1.4 kpc × 5.6 kpc, deprojected) field to have a side length several times the expected gas scale height (e.g., M. Heyer & T. M. Dame 2015; K. Yim et al. 2011, 2014; S. M. R. Jeffreson et al. 2022) and span spiral arm and interarm regions. The data have full flux recovery and a high enough physical resolution (7 pc) to see small bound objects in the context of their kiloparsec-scale superstructures.

In Section 2, we describe these observations and our data reduction. In Section 3, we apply a multiscale dendrogram decomposition algorithm to unpack these data and estimate the key physical properties of CO-emitting structures across multiple spatial scales. Our primary results are presented in Section 4, in which we explore the scaling of $\alpha_{\rm vir}$ against physical and hierarchical parameters as well as complementary scaling relations. In Section 5 we present a synthetic picture of our results and discuss some caveats. Finally, we summarize these results and conclude in Section 6.

## 2. Data: Zooming in on NGC 253

### 2.1. Observations with ALMA

We obtained ALMA Band 6 data to capture the $^{12}$CO, $^{13}$CO, and C$^{18}$O($J = 2$–1) lines over a region in NGC 253's disk (Figure 1; projects #2017.1.01101.S and #2018.1.00596.S,

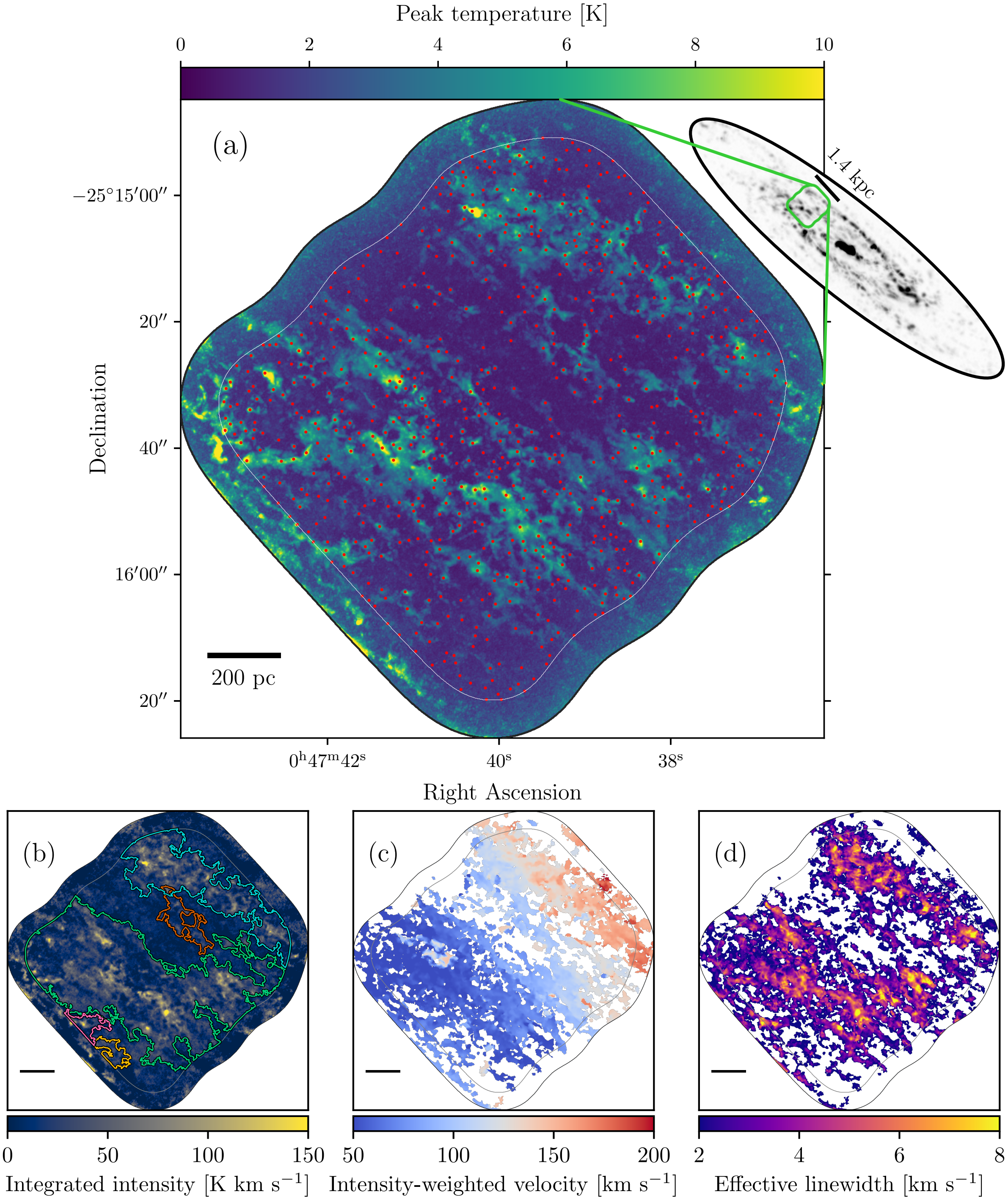

**Figure 1.** (a) Peak $^{12}$CO(2–1) temperature of the 1.4 × 5.7 kpc$^2$ (deprojected) field along each line-of-sight pixel, with a spatial resolution of ∼7 pc. The peak locations of dendrogram leaves are plotted as red points. Where leaf footprints overlap, we show only the brightest. The 6″ border is shown in white. In the upper right, an ACA map shows the full-galaxy context in grayscale. (b) Integrated intensity (zeroth moment) of the field. Five major trunks of the dendrogram, those containing eight or more hierarchical levels, are highlighted as contours in projection using the same colors as in Figure 2. (c) Intensity-weighted velocity (first moment) of the field using a high-sensitivity "strict" mask. (d) Effective linewidth of the field using the strict mask.

PI: E. Rosolowsky). We combined observations using two 12 m configurations with the 7 m and total power (TP) arrays. As a result, our observations resolve all spatial scales greater than the 0″.4 (∼7 pc) synthesized beam over a field of 80″ × 80″(∼1.4 kpc × 5.6 kpc, deprojected). The observations include a total on-source time of 1374.91 minutes (TP), 486.19 minutes (7 m), and 149.68 minutes (12 m), with a mean elevation at 54°–83° and PWV between 0.4 and 1.8 mm, reflecting near-optimal conditions. This work focuses on the $^{12}$CO(2–1) line, and isotopologues will be analyzed in a future study.

### 2.2. Data Reduction and Imaging

We processed and imaged these data via the PHANGS-ALMA pipeline, which is described in detail by A. K. Leroy et al. (2021a). For the 12 m and 7 m observations, the pipeline

**Table 1**
$^{12}$CO(2–1) Cube Properties

| Quantity | Value |
|---|---|
| Arrays | 12m (C43-5, C43-2) +7m + TP |
| Resolution | |
| Angular [arcsec] | 0.42 |
| Physical [pc] | 7.1 |
| Channel width [km s$^{-1}$] | 0.95 |
| Pixel scale [arcsec] | 0.084 |
| Area mapped | |
| Angular [arcmin$^2$] | 1.8 |
| Deprojected physical [kpc$^2$] | 7.6 |
| 1$\sigma$ noise[a] | |
| [mJy beam$^{-1}$] | 3.1 |
| [mK] | 400 |
| [$M_\odot$][b,c] | 300 |
| 1$\sigma$ surface brightness noise[b] | |
| [K km s$^{-1}$] | 0.89 |
| [$M_\odot$ pc$^{-2}$][c] | 5.3 |
| Spectral completeness[d] [%] | 94 |

**Notes.**
[a] Median of the noise cube within the 6″ border.
[b] For line-integrated mass and surface brightness sensitivities, we adopt a linewidth of $\Delta v = 5$ km s$^{-1}$. This implies an improvement of $\sim\sqrt{5}$ in the surface brightness sensitivity.
[c] For masses, we adopt a constant $\alpha_{\rm CO}^{1-0}$ and $R_{21}$ (see Equation (4) and Appendix B).
[d] Defined here as `SUM(ACA_masked)/SUM(ACA)` (see Section 2.3).

begins by extracting and staging the calibrated $u - v$ data and subtracting the continuum with a fit that excludes significant emission lines. After continuum subtraction, a line-specific data set is constructed by regridding and recombining the data to a common velocity grid. The calibrated measurement sets are then imaged by repeated calls to `CASA`'s `tclean` task (CASA Team et al. 2022), using a mixture of multiscale and single-scale clean calls.

For the reduction of TP data, the pipeline involves a modified version of the procedure described in C. N. Herrera et al. (2020), including calibration, baseline fitting, unit conversion and data concatenation, and imaging. In our case, however, limitations of the Band 6 intermediate frequency (IF) gap at the time of observations precluded automatically setting a line-free velocity range to baseline. Thus, we manually reprocessed the TP data to fit a linear baseline, using emission-free regions in the TP cube along with a carefully defined manual mask.

After imaging and deconvolving the 12 m+7 m and TP cubes, the pipeline applies a primary beam correction and convolves the elliptical synthesized beams to round ones. We then construct a continuous field via linear mosaicking. The cleaned 12 m+7 m cube is combined with the single-dish TP cube by aligning and feathering in the Fourier-transformed domain. The data are then downsampled by rebinning pixels to minimize the data volume while ensuring the beam is well-sampled. Finally, a unit conversion from Jy beam$^{-1}$ to kelvin brings our cube into a science-ready state.

In parallel, we additionally constructed a three-dimensional noise cube to estimate the rms noise in our data, following the procedure outlined in A. K. Leroy et al. (2021a).

The resulting $^{12}$CO(2–1) data and noise cubes are described in Table 1. There, we note the array combination, resolution, area mapped, noise, surface brightness noise, and completeness. The data have an angular resolution of 0$''$.42 × 0$''$.42 and a spectral resolution of 0.95 km s$^{-1}$ over an area of 1$'$.8$^2$. Using a distance of 3.5 Mpc (S. Okamoto et al. 2024; M. J. B. Newman et al. 2024) and an inclination of 76° (A. McCormick et al. 2013), this allows us to resolve and characterize ∼3–300 pc-scale (deconvolved $R$) structures over a deprojected field of 7.6 kpc$^2$.

### 2.3. *Sensitivity and Completeness of the Cube*

To remove noisy regions at the edge of our field, we applied a 6″ mask around the image borders (which represents the distance where the median noise exceeds the 75th percentile of the full distribution) and report the median noise within the inner, stable noise region in Table 1. When determining line-integrated surface brightness sensitivities, we assume a fiducial linewidth of 5 km s$^{-1}$. To convert sensitivities to physical mass units, we estimate a constant $\alpha_{\rm CO}^{(1-0)} = 3.68\,M_\odot$ (pc$^2$ K km s$^{-1}$)$^{-1}$ and $R_{21} = 0.615$, as discussed in Appendix B. Our typical 1$\sigma$ sensitivity of 400 mK implies a point-mass sensitivity of 300 $M_\odot$ within the beam and a 1$\sigma$ mass surface density sensitivity of $\Sigma_{\rm mol} = 5.3\,M_\odot\,{\rm pc}^{-2}$. Typical resolved GMCs have masses $\sim10^4$–$10^5\,M_\odot$ and surface densities of a few times 10–100 $M_\odot$ pc$^{-2}$ (E. Schinnerer & A. K. Leroy 2024), which suggests that our map is sensitive to sub-cloud-scale structures.

To estimate signal completeness, we construct two masks of CO emission in the cube: a high-confidence "strict" mask and a high-completeness "broad" mask. Following E. Rosolowsky & A. Leroy (2006) and A. K. Leroy et al. (2021a), the strict mask is constructed using regions where the emission exceeds a signal-to-noise ratio of 4 over two consecutive velocity channels, subsequently extended to adjacent regions with a signal-to-noise ratio exceeding 2. Strict masks constructed across several resolutions, including coarse resolution where averaging improves the signal-to-noise, are merged to define the broad mask. The broad mask includes nearly all emission in the cube.

Due to bandwidth limitations (Section 2.2), the $^{12}$CO(2–1) flux in our cube is not fully captured by the spectral range that we have surveyed. To estimate the flux outside the interferometer's spectral response, we use ACA (7 m + TP) observations of NGC 253 from project #2018.1.01321.S (A. K. Leroy et al. 2021b, PI: C. Faesi) reduced with the PHANGS-ALMA pipeline as in Section 2.2, which fully covers the CO emission in the spectral domain. We then mask the ACA cube twice: first to the projected area of our field, and then to its corresponding spectral range. By comparing the total ACA flux within this mask to the total flux across all velocities, we find that our data sample 94% of the significant emission from the full spectral range (Table 1). We thus consider the flux to be adequately sampled for our needs.

### 2.4. *Additional Data Products*

In Figure 1, we present maps of (a) the peak temperature, (b) unmasked line-integrated intensity (zeroth moment), (c) strictly masked intensity-weighted mean velocity (first moment), and (d) strictly masked effective linewidth of the $^{12}$CO(2–1) cube, following the definition of M. H. Heyer et al. (2001). In panel (a), the peak temperature locations of leaves in the dendrogram identified in Section 3.1 are plotted as red points, and in panel (b) we highlight the projected outlines of five major dendrogram trunks similarly marked in Figure 2. These data products will be made available online.

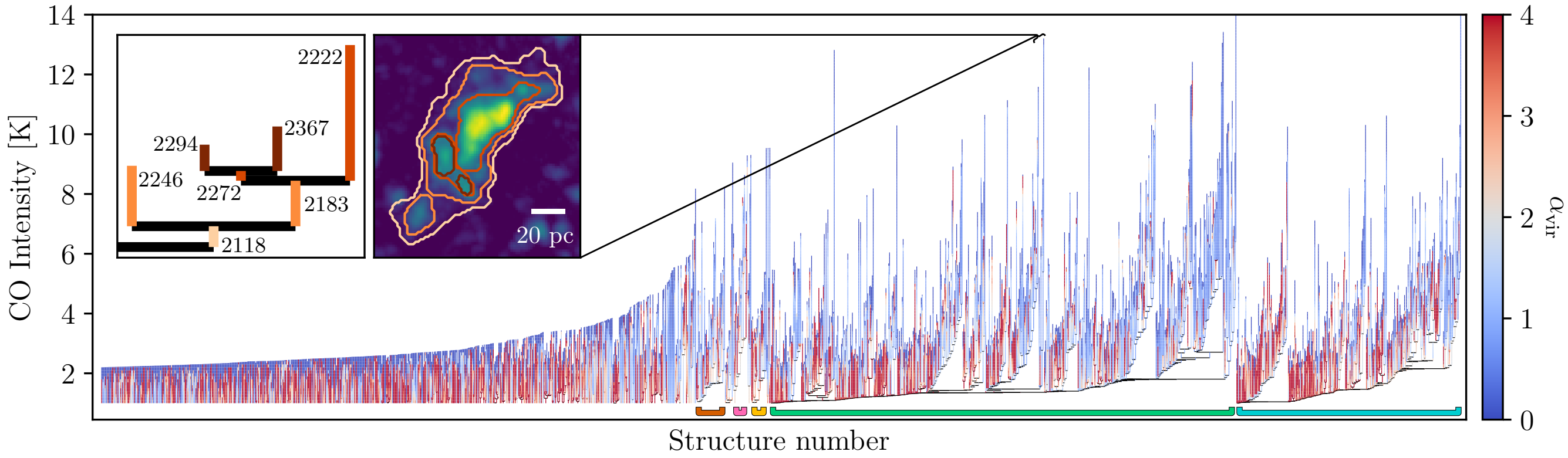

**Figure 2.** Main panel: full dendrogram of the NGC 253 field, showing 2463 structures ordered by peak intensity within a lineage. The virial parameter $\alpha_{vir}$ is plotted within each structure as a function of the contour intensity level, showing a substantial decrease with intensity both between (Section 4.2.3) and within (Section 4.4.1) individual structures. The trunks corresponding to base structures #1018 (orange), 1560 (pink), 1439 (yellow), 488 (green), and 87 (cyan) are indicated at the bottom, using the same colors as Figure 1, panel (b). Inset panels: a subset of the dendrogram (left) and corresponding contours over the peak intensity map (right) highlight the tree-like hierarchy of the dendrogram, in which leaves (the high-intensity, peak-level structures) are nested within lower-intensity parent branches, which can in turn be nested within their own parent structures. In these panels, the orange gradient represents each structure's level.

The 6″ border, which excludes higher-noise regions at the edge of our map, is shown by a thin white (a) or gray (b)–(d) line. Subsequent analyses in this work use data within the 6″ border but without a mask for noise or completeness.

## 3. Hierarchical Dendrogram Analysis

### 3.1. Structural Decomposition

Since we are interested in identifying structure at all spatial scales, we use the `astrodendro`[26] (T. Robitaille et al. 2019) implementation of E. W. Rosolowsky et al.'s (2008) hierarchical dendrogram algorithm. Detailed descriptions of the dendrogram algorithm are given in E. W. Rosolowsky et al. (2008), A. A. Goodman et al. (2009), and R. Shetty et al. (2012). In brief, `astrodendro` constructs a tree-like hierarchy by starting with the highest-intensity voxels and progressively adding fainter ones. Each local maximum above a given intensity threshold defines a new "leaf". At the adjacent boundaries of neighboring structures, these top-level leaves are joined into a "branch" which combines their boundaries; this process is continued, merging new branches as needed, until a minimum intensity level is reached. Each voxel can therefore be assigned to only a single leaf but to multiple nested branches within a given tree.

Three parameters tune the execution of this algorithm:

1. The minimum value (`min_value`) sets the intensity limit below which no voxels are added to the dendrogram. We choose a value of 1.0 K, corresponding to a 2.5$\sigma$ noise threshold (Table 1).
2. The minimum significance between adjacent structures (`min_delta`) determines the minimum spacing between isocontours so that low-significance local maxima are not immediately assigned to new structures. Here, we choose a threshold of 1.2 K or 3$\sigma$.
3. Finally, we choose the minimum number of voxels in a structure (`min_npix`) to be four times the beam solid angle $\Omega$,

$$N = \frac{4\pi\theta^2}{4\ln(2)A} = 116 \text{ voxels}, \tag{1}$$

where $\theta$ is the full width at half maximum (FWHM) assuming a Gaussian beam and $A$ is the projected area per pixel. We add one factor of 2 to include only spatially resolved structures, and an additional factor of 2 to account for the three-dimensional extent of structures in phase space.

We characterize the impact of these parameter choices on our dendrogram and science results in Appendix C, finding no significant effect on our conclusions for a reasonable range of choices.

The fiducial dendrogram decomposition of our field, with 1500 substructure-free leaves and 963 branches, is shown in Figure 2. Inset panels highlight a subset of the dendrogram and the corresponding projected contours, colored by hierarchical level. Using the labeled structures in this inset as a reference, we clarify some dendrogram terminology below:

*Structure.* A closed PPV contour in the dendrogram, i.e., a leaf or a branch, which includes the substructures within it.

*Leaf.* A peak-level structure, either isolated or at the top of the hierarchy (#2247, 2294, 2367, and 2222).

*Branch.* A structure that contains substructure in the form of additional leaves or branches (#2118, 2183, and 2272).

*Parent.* The directly preceding structure in the hierarchy (here, #2118 is the "parent" structure of #2246 and 2183, which are therefore its "children").

*Lineage.* The entire descendant hierarchy of a given structure (the inset shows the lineage of branch #2118, which is part of the lineage of branch #488).

*Root.* A branch at level 0 (#488 for the example given here; not shown in the inset but highlighted with green in the main figure).

The dendrogram of our field shows a high degree of structure, with up to 176 hierarchical levels. For leaves with at least one parent, the median level (number of hierarchical steps from the lowest-intensity root) is $61^{+58}_{-58}$, where uncertainties represent the 16%–84% range. In Figure 1, we plot the locations of leaves as red points. Two major roots (#488 and 87) and their descendants dominate the highly structured regime and correspond to the two arm-like features of NGC 253 that intersect our field, extending along the northeast–southwest axis. Figure 1 shows the outer boundaries of these and three other roots with at least eight descendant levels (#1018, 1560, and 1439). These large complexes are identified in Figure 2 by the colored, horizontal bars below their dendrogram trees. We defer discussion of the virial parameters shown in this figure to the results.

[26] http://www.dendrograms.org/

### 3.2. Estimating Physical Properties

For each structure, we derive the radius, velocity dispersion, mass, surface density, and virial parameter, following the moment methods of E. Rosolowsky & A. Leroy (2006) and E. W. Rosolowsky et al. (2008). We correct for sensitivity and resolution biases by extrapolating and deconvolving our measured masses, radii, and velocity dispersions. Our full structure catalog, including derived parameters and bootstrapped uncertainties, will be made available online in machine-readable format.

#### 3.2.1. Radius

Following E. W. Rosolowsky et al. (2008), we define the rms size of a structure $\sigma_r$ as the geometric mean of the semimajor and semiminor axes such that

$$\sigma_r = \sqrt{\sigma_{\rm min}\sigma_{\rm maj}}. \tag{2}$$

$\sigma_{\rm min}$ and $\sigma_{\rm maj}$ are, in turn, computed from the intensity-weighted second moment along each axis of the structure's projection in the position–position plane. As in E. Rosolowsky et al. (2021), we model the surface brightness of our structures as a two-dimensional Gaussian. To convert the one-dimensional rms size $\sigma_r$ to the appropriate half width at half maximum radius

$$R = \eta\sigma_r, \tag{3}$$

we therefore apply a factor of $\eta = \sqrt{2\ln(2)} \approx 1.18$. This conversion factor is smaller than the factor of 1.91 empirically derived by P. M. Solomon et al. (1987) and widely used since then, which should be noted for the purposes of any comparison. The spherical approximation used here breaks down when the structure size $R$ approaches the galaxy scale height, but our structures are almost all below this threshold and do not exhibit a significant preference toward in-plane alignment (Section 3.2.7 and Appendix D).

#### 3.2.2. Velocity Dispersion

We define the velocity dispersion $\sigma_{\rm v}$ as the intensity-weighted second moment of velocity for a given structure, which is related to the linewidth by a linear factor for a Gaussian distribution. N. Krieger et al. (2020) confirmed that several different definitions of velocity dispersion can shift scaling relations by a constant factor but do not substantially change the slope.

#### 3.2.3. Luminosity and Mass

We derive a CO-based mass estimate using a conversion from the structure's luminosity. The flux of a structure is the sum over the velocity-integrated intensity of the emission in the region. To get the luminosity, we scale this flux by the area of each pixel in square parsecs. To convert from luminosity to mass, we apply a constant CO-to-$H_2$ conversion factor $\alpha_{\rm CO}$ and line ratio $R_{21}$ such that

$$M_{\rm CO} = \alpha_{\rm CO}^{(2-1)} L = \frac{\alpha_{\rm CO}^{(1-0)}}{R_{21}} L. \tag{4}$$

Given the established variations in $\alpha_{\rm CO}$ and $R_{21}$ with respect to environment, we explore several prescriptions for each of these parameters in Appendix B. Ultimately, we choose to adopt a metallicity-dependent $\alpha_{\rm CO}$ based on E. Schinnerer & A. K. Leroy (2024), finding $\alpha_{\rm CO}^{(1-0)} = 3.68\ M_\odot$ (pc$^2$ K km s$^{-1}$)$^{-1}$ for a value of $\langle Z\rangle = 1.12\ Z_\odot$ (Equation (B1)). For $R_{21}$, we use data from the Nobeyama 45 m telescope (K. Sorai et al. 2000; N. Kuno et al. 2007) and ACA observations (A. K. Leroy et al. 2021b) to directly measure $R_{21} = 0.615 \pm 0.180$. These combine to give $\alpha_{\rm CO}^{(2-1)} = 5.99\ M_\odot$ (pc$^2$ K km s$^{-1}$)$^{-1}$, and based on the range of $\alpha_{\rm CO}$ prescriptions impose a systematic uncertainty of 35% on our mass measurements. As $R_{21}$ and $\alpha_{\rm CO}$ are calibrated on significantly larger scales, local differences in the gas physical conditions may drive significant variations on scales of a few parsecs (e.g., M.-Y. Lee et al. 2014).

#### 3.2.4. Surface Density

We define the surface density for a structure as

$$\Sigma_{\rm mol} = \frac{M_{\rm CO}}{2\pi R^2}, \tag{5}$$

representing the average surface density within the FWHM radius for a two-dimensional Gaussian structure (E. Rosolowsky et al. 2021).

#### 3.2.5. Virial Parameter and $R_{vir}$

To investigate the energy balance and dynamical state of structures, we estimate the virial parameter as

$$\alpha_{\rm vir} \equiv \frac{2K}{U_{\rm grav}} = \frac{2M_{\rm vir}}{M_{\rm CO}} = \frac{10\sigma_{\rm v}^2 R}{fGM_{\rm CO}}, \tag{6}$$

where $f$ is a geometrical factor that describes the assumed mass density profile $\rho(r)$ (P. M. Solomon et al. 1987; F. Bertoldi & C. F. McKee 1992; C. F. McKee & E. G. Zweibel 1992). For a power-law radial profile $\rho(r) \propto r^{-k}$, $f = \frac{1-k/3}{1-2k/5}$. Following E. Rosolowsky & A. Leroy (2006), we assume $k = 1$ such that $f = 10/9$. The second factor of 2 arises because the appropriate dynamical comparison is between $M_{\rm vir}$ and the luminous mass contained within the FWHM of our Gaussian cloud model, $M_{\rm CO}/2$ (E. Rosolowsky et al. 2021; J. Sun et al. 2022).

As we are also interested in $\alpha_{\rm vir}$'s change along the structure hierarchy, we define the ratio

$$R_{\rm vir} = \frac{\alpha_{\rm child}}{\alpha_{\rm parent}}, \tag{7}$$

which compares the higher-intensity, inner child structure to its lower-intensity, directly preceding parent branch, where applicable (Figure 2, inset). This implies that structures with $R_{\rm vir} < 1$ have lower $\alpha_{\rm vir}$ and are thus more gravitationally bound relative to their enveloping parent structures.

In the absence of other forces, $\alpha_{\rm vir}$ characterizes the relative strength of gravitational binding energy and the kinetic energy of a structure. Assuming that velocity motions are mostly supportive, the threshold for a hypothetical, long-lived object to be virialized or "bound" is therefore $\alpha_{\rm vir} = 2$. However, we emphasize that this should not be overinterpreted as a physical threshold for boundedness, which would require long-lived, well-defined objects under the influence of gravity and their kinetic energy alone. Observationally, this scenario is further obscured by uncertainties in $\alpha_{\rm CO}$, $f$, and other systematics discussed in Section 5.3.

#### 3.2.6. Bias Corrections and Statistical Uncertainties

We correct for sensitivity bias and clipping of the dendrogram structures by extrapolating measured properties to those we would measure at a 0 K contour, i.e., for observations of perfect sensitivity. To determine the appropriate extrapolation approach, we study the profiles of spatial moments, the velocity moment, and flux as a function of sensitivity for individual structures (e.g., E. Rosolowsky & A. Leroy 2006, their Figure 2). As in that work, we find that linear extrapolations are appropriate for the spatial and velocity moments ($\sigma_{\rm x}$, $\sigma_{\rm y}$, and $\sigma_{\rm v}$). However, the quadratic flux extrapolation that they recommend can lead to unstable results for dendrogram structures, since the minimum sensitivity threshold for high-level structures can be significantly above the 0 K floor. We therefore follow the approach presented in E. Rosolowsky & L. Blitz (2005) and A. K. Leroy et al. (2015) and apply a Gaussian correction to our fluxes, which is more consistent with the observed flux profiles. After this correction, leaves comprise 57% of the total flux in our field.

We further correct for resolution bias by deconvolving the spatial beam and channel width from the measured radius and velocity dispersion, respectively.

We use bootstrapping to estimate statistical uncertainties in the structural properties, following E. Rosolowsky & A. Leroy (2006). For each measured structure, this involves generating a sample of $N$ trial structures consisting of ($x$, $y$, $v$, $T$) data randomly drawn from its distribution. We then calculate the structure properties for each of these iterations. To test how many bootstrap iterations are needed to achieve convergence in the uncertainties, we studied the scaling for a subset of 15 randomly selected leaves and 15 randomly selected branches, settling on $N = 200$ as a reliable threshold. Of course, bootstrapping only reflects the uncertainty in determining structure properties once the dendrogram has been defined and does not capture methodological challenges involved in decomposing the data itself leading to systematic biases.

#### 3.2.7. Inclination Corrections

At $\sim$100 pc scales, pixel-by-pixel measurements of J. Sun et al. (2020) and CPROPS-identified structures of Hughes et al. (2025, in preparation, private communication) have revealed correlations between the inclination angle of host galaxies and observed properties of the molecular gas in the PHANGS-ALMA sample. The empirically derived corrections of J. Sun et al. (2022) for $\Sigma_{\rm mol}$ and $\sigma_{\rm v}$ would introduce an additional factor on $\alpha_{\rm vir}$ of 1.27 for large structures, where $R$ approaches the galaxy scale height, and 1 (i.e., no change) for small structures (Equation (6)).

However, we do not apply inclination corrections for the following reasons:

1. *Small structures relative to scale height.* Inclination and shear effects are likely subdominant for structures smaller than the galaxy's scale height (e.g., Y.-H. Xie et al. 2024; Y.-H. Xie & G.-X. Li 2025). To approximately estimate NGC 253's scale height in our region of interest, we adopt inclination-corrected values of the region-averaged stellar mass and molecular gas surface densities $\langle \Sigma_\star \rangle = 372\,M_\odot\,{\rm pc}^{-2}$ and $\langle \Sigma_{\rm mol} \rangle = 26\,M_\odot\,{\rm pc}^{-2}$ from the "mega-tables" catalog of J. Sun et al. (2022) and J. Sun et al. (2023). Under the assumption of weakly self-gravitating gas, $h \sim \sigma_{\rm v,z}(2\pi {\rm G}\Sigma_\star / h_\star)^{-1/2}$, while a gas-dominated potential gives $h \sim \sigma_{\rm v,z}^2 (8\pi)^{-1/2} ({\rm G}\Sigma_{\rm mol})^{-1}$ (e.g., H. Koyama & E. C. Ostriker 2009). For a typical $\sigma_{\rm v,z} = 5\,{\rm km\,s^{-1}}$ and stellar scale height $h_\star = 500$ pc, $h_{\rm FWHM} \sim 100$ pc in either case. This first-order estimate is in agreement with observations of the Milky Way (M. Heyer & T. M. Dame 2015; J. Roman-Duval et al. 2016; A. Marasco et al. 2017), nearby galaxies (K. Yim et al. 2011, 2014; C. Bacchini et al. 2019), and targeted simulations (S. M. Benincasa et al. 2016; C.-G. Kim & E. C. Ostriker 2017; S. M. R. Jeffreson et al. 2022), which find FWHM molecular gas scale heights of 50–200 pc or greater at intermediate distances in the disk. In our catalog, the largest leaf radius is just 32 pc and 74% of branches (89% of all structures) are below $R = 100$ pc, suggesting they are likely too small for an inclination correction to be appropriate.
2. *Near-randomly aligned aspect ratios.* In Appendix D, we show that structure aspect ratios are not preferentially aligned with the major axis of NGC 253 to a significant degree.
3. *Methodological uncertainty.* Finally, it is unclear how the corrections of J. Sun et al. (2022), derived at $\sim$100 pc scales, apply to our much higher-resolution data. Furthermore, since we are particularly interested in the *gradient* of $\alpha_{\rm vir}$ with quantities such as $R$ and $\Sigma_{\rm mol}$, a scale-dependent inclination correction could give rise to spurious (apparent) characteristic scales.

### 3.3. Completeness Limits

We assess the completeness of our catalog by injecting simulated Gaussian sources into signal-free portions of the $^{12}$CO(2–1) data cube, following the approach presented in E. Rosolowsky et al. (2021). To define a signal-free region, we combine the 6″ border mask with the broad emission mask described in Section 2.3 and expand it by the three-dimensional FWHM radius of the beam, inverting the resulting mask. 49% of all voxels are included in the signal-free cube.

The properties of our simulated sources are drawn from log-uniform distributions of mass, surface density, and virial parameter, adopting a fixed $\alpha_{\rm CO}$ as in Section 3.2.3. Masses are drawn from a 2.5 dex distribution centered around $M_{\rm CO} = 7600\,M_\odot$, corresponding to a 50$\sigma$ detection in a single beam and single channel. Surface densities are drawn from a 2.5 dex distribution centered around $\Sigma_{\rm mol} = 150\,M_\odot\,{\rm pc}^{-2}$. Virial parameters are drawn from a 2 dex distribution centered around $\alpha_{\rm vir} = 2$. We use these values with Equations (5) and

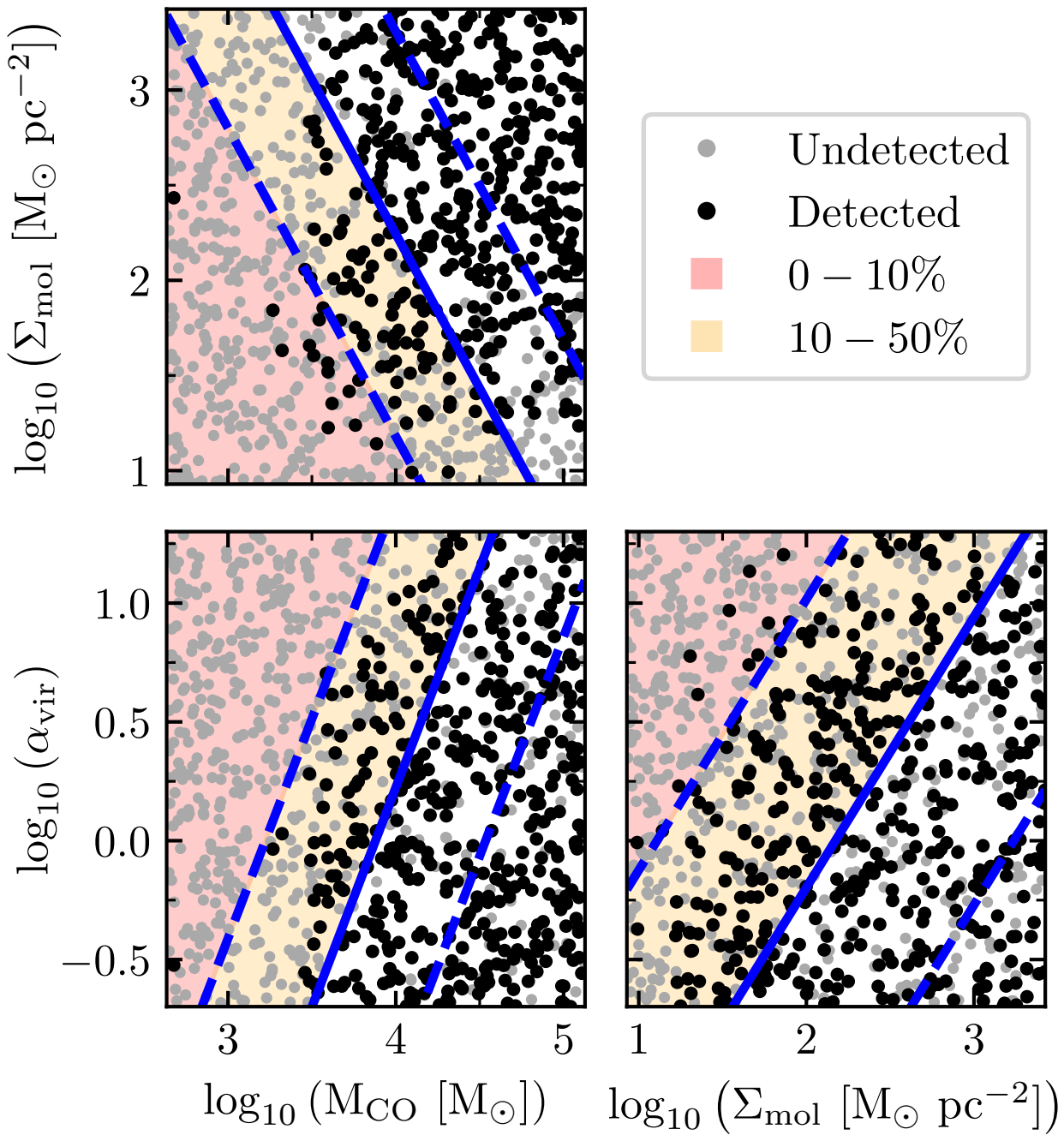

**Figure 3.** Recovery of 1005 test sources injected into signal-free regions of the $^{12}$CO(2–1) cube. We plot the injected $\Sigma_{\rm mol}$, $\alpha_{\rm vir}$, and $M_{\rm CO}$ against each other and indicate whether or not the corresponding source is detected. In each panel, the solid blue line shows a logistic regression to determine where structures would be detected with 50% completeness, while the neighboring dashed lines show analogous boundaries of 10% and 90%.

(6) to calculate the implied two-dimensional radius $R$ and the velocity dispersion $\sigma_{\rm v}$ for a Gaussian structure.

We inject $\sim 10^3$ of these false sources into the signal-free portion of the data cube and follow the same approach of our main analysis to produce a dendrogram catalog. For each source, we note whether a leaf was detected at its central coordinate. To determine the overall probability of detecting a leaf with a given $M_{\rm CO}$, $\Sigma_{\rm mol}$, and $\alpha_{\rm vir}$, we fit a logistic regression to the detection fraction using the STATSMODELS (S. Seabold & J. Perktold 2010) package and E. Rosolowsky et al.'s (2021) Equation 13.

The 50% completeness level for this model is shown by the solid blue lines of Figure 3. Roughly half the injected sources near these lines are detected, demonstrating that the regression is a fair approximation to the completeness of our data. Qualitatively, the figure shows that $M_{\rm complete}$ increases with decreasing $\Sigma_{\rm mol}$ and with increasing $\alpha_{\rm vir}$. Structures with very low surface densities ($\Sigma_{\rm mol} \lesssim 20\,M_\odot\,{\rm pc}^{-2}$) and/or very large virial parameters ($\alpha_{\rm vir} \gtrsim 20$) are poorly recovered regardless of their mass, reflecting the effect of spreading a fixed signal out over a wide spatial and spectral range. Similarly, we miss low-mass ($M_{\rm CO} \lesssim 10^3\,M_\odot$) sources regardless of $\alpha_{\rm vir}$ and $\Sigma_{\rm mol}$, which biases us to higher-mass structures in the completeness-limited regime.

For fiducial values of the surface density and virial parameter ($\Sigma_{\rm mol} = 150\,M_\odot\,{\rm pc}^{-2}$ and $\alpha_{\rm vir} = 2$) the 50% completeness level in mass is $M_{\rm complete} = 1.1 \times 10^4\,M_\odot$, implying a corresponding $R_{\rm complete} = 3.4$ pc and $\sigma_{\rm v,complete} = 1.7\,{\rm km\,s^{-1}}$. Looking at the detection fraction as a function of $\Sigma_{\rm mol}$ and $\alpha_{\rm vir}$, we find the smoothed distribution exceeds 50% above a value of $\Sigma_{\rm mol,complete} = 59\,M_\odot\,{\rm pc}^{-2}$ and below $\alpha_{\rm vir,complete} = 2.4$.

## 4. Results

In the following sections, we explore the physical properties of the dendrogram structures with a particular focus on the scaling of $\alpha_{\rm vir}$. We use extrapolated and deconvolved values throughout, except where otherwise indicated.

Where linear fits are appropriate, we use the Bayesian Markov Chain Monte Carlo (MCMC) linear regression routine `linmix`[27] fit to the centers of binned data (B. C. Kelly 2007). We exclude the bins below our estimated 50% completeness limits as well as bins dominated by a high-radius, high-mass lineage discussed further in Section 5.2. For the uncertainty in the $x$- and $y$-axis quantities, we take half the bin width and the standard deviation divided by the square root of the bin population, respectively. We estimate uncertainties on fit slopes using the standard deviation of a posterior distribution with 10,000 draws. To control for the effects of binning on our results, we vary the number of bins by $\pm 20\%$, which yields only minor changes in fitted slopes ($<2\sigma$) and does not affect any qualitative interpretations.

### 4.1. The Virial Parameters of Multiscale Structures

The distribution of the virial parameter is shown in the right panel of Figure 4 and described in Table 2. Evidently, leaves and branches have very consistent $\alpha_{\rm vir}$ as a population, something that is true for all binned sizes. Leaves are marginally but significantly more bound than branches, with a 95% uncertainty on the mean of 0.1. Figure 4 and Table 2 show that $\alpha_{\rm vir}$ is much better constrained for branches than for leaves. The observed scatter reflects uncertainty in our calculations as well as intrinsic variations in the physical state of objects. This trend is mostly driven by structure sizes, as smaller structures—typically leaves—tend to have fewer voxels and hence more scatter in the bootstrapped properties. Of course, bootstrapped uncertainties do not fully capture the systematic errors that influence our calculations, which are reviewed in Section 5.3, in addition to the 35% mass uncertainty introduced by $\alpha_{\rm CO}$ and $R_{21}$ (Section Appendix B).

In order to produce the star formation efficiencies of a few percent or less that are observed on cloud (50–100 pc) scales (A. K. Leroy et al. 2025), theoretical models invoking turbulence and/or stellar feedback predict that most clouds should be unbound and exhibit relatively high virial parameters ($\alpha_{\rm vir} \gtrsim 4$; see, e.g., P. Padoan et al. 2012; V. A. Semenov et al. 2017; N. J. Evans et al. 2022; H. Fukushima & H. Yajima 2022), even when a wide range of feedback processes are considered (J.-G. Kim et al. 2021). Our results, however, suggest that typical $\sim$3–250 pc-scale structures in the field are marginally bound with $\alpha_{\rm vir} \approx 2$. This hews more closely to observational surveys such as the corrected $\alpha_{\rm vir}$ of L. Liu et al. (2021) or the aperture-based results of J. Sun et al. (2022) as well as other dendrogram decompositions of CO fields such as T. S. Rice et al. (2016)

The gravitational binding energy for elongated structures can change depending on the aspect ratio (F. Bertoldi & C. F. McKee 1992). As shown in Appendix D, the majority of structures in our catalog are not highly extended or filamentary; nevertheless, it is important to consider how appropriate corrections might affect our results. Taking into account ellipsoidal corrections does bring our measured

[27] https://linmix.readthedocs.io/en/latest/index.html

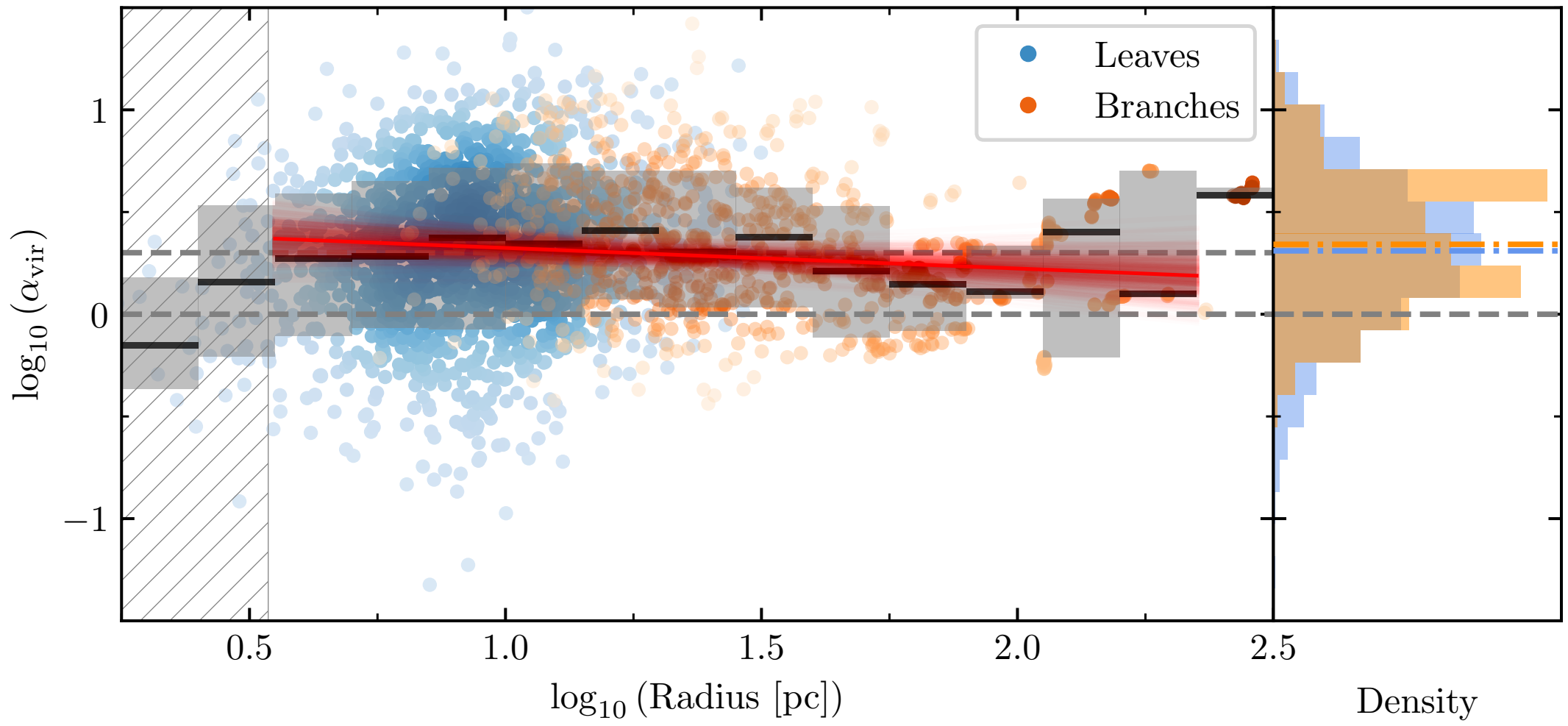

**Figure 4.** Left: the virial parameter against structure radius for all leaves (blue) and branches (orange) in the dendrogram. Each point's lightness is scaled by a Gaussian kernel density estimator (KDE) to reflect the smoothed density distribution. Bins of 0.15 dex are shown with their medians in black and 16%–84% ranges in gray. In red, we fit the centers of these bins and show each iteration of a Bayesian linear regression using `linmix`, finding a slope of $-0.10 \pm 0.08$. We exclude the largest-radius bin because it is strongly affected by a single, very large lineage with duplicate flux, discussed more in Section 5.2. The two dashed, horizontal lines in gray show $\alpha_{\rm vir} = 1$ and 2. The estimated 50% completeness limit for structures of typical $\Sigma_{\rm mol}$ and $\alpha_{\rm vir}$ (Section 3.3) is shown by the hatched boundary. $\alpha_{\rm vir}$ is remarkably flat with $R$, for structures from ∼3 to 250 pc. Right: virial parameter distributions for leaves (blue) and branches (orange). Median values are indicated by the horizontal, colored lines.

**Table 2**
Median $\alpha_{\rm vir}$, Fractional Bootstrapped Errors, and Bound Fraction for Leaves ($N = 1500$), Branches ($N = 943$), and the Full Structure Catalog ($N = 2463$)

| Quantity | Leaves | Branches | All |
|---|---|---|---|
| $\alpha_{\rm vir}$ | $2.0^{+2.8}_{-1.2}$ | $2.2^{+2.0}_{-1.1}$ | $2.1^{+2.4}_{-1.2}$ |
| $\lvert\delta\alpha_{\rm vir}/\alpha_{\rm vir}\rvert$ | $1.1^{+0.8}_{-0.5}$ | $0.1^{+0.3}_{-0.1}$ | $0.7^{+0.9}_{-0.6}$ |
| $\alpha_{\rm vir} < 2$ | 48% | 46% | 47% |

**Note.** Uncertainties represent the 16%–84% range of values.

$\alpha_{\rm vir}$ closer to, but still much lower than, the high virial parameters discussed above–when we apply the corrections found in F. Bertoldi & C. F. McKee (1992; their Appendix A) to estimate ellipsoidal virial parameters, the resulting distribution is uniformly higher (median factor of 1.3). However, the shape of the scaling relations presented in this section does not change, and the fitted slopes are consistent to our quoted uncertainties. Similarly, we may consider a filamentary stability criteria in which the critical virial mass per unit length is

$$(M/L)_{\rm crit} = 2\sigma_{\rm v}/{\rm G} \tag{8}$$

(J. D. Fiege & R. E. Pudritz 2000; G.-X. Li et al. 2016). Applying this criteria to our most elongated structures (aspect ratio >3), we observe no statistically significant changes to the correlations discussed below.

### 4.2. Self-gravity and Its Scaling

#### 4.2.1. Scaling with Radius

In the left panel of Figure 4, we plot the virial parameter for individual leaves (blue) and branches (orange) against their radius. In general, we find no evidence that gravitational boundedness changes with spatial scale. The median $\alpha_{\rm vir}$ remains remarkably invariant across 2 orders of magnitude in scale, which is demonstrated by a linear fit consistent to $<1.5\sigma$ with zero. The smallest and largest bins offer exceptions to this flat behavior, but this is explained by other effects: structures below our completeness limit are biased toward lower $\alpha_{\rm vir}$ (Section 3.3), while $\alpha_{\rm vir}$ estimates for the largest structures are contaminated due to large-scale motions (Section 5.2).

Previous studies have reported varying $\alpha_{\rm vir}$–$R$ scaling exponents. Numerical simulations by R. Shetty et al. (2010) found anticorrelations ($\alpha_{\rm vir} \propto R^{-1.1}$ in 3D space; $\alpha_{\rm vir} \propto R^{-0.8}$ in PPV space), attributing this to pressure confinement effects. Similarly, M. R. Krumholz et al. (2019) proposed $\alpha_{\rm vir} \propto R^{-1}$ for clouds following the size–linewidth relation, based on turbulent regulation theories. However, T. J. O'Neill et al. (2022) demonstrated that $\alpha_{\rm vir}$–$R$ scaling depends significantly on density profiles and size–linewidth indices. Our observed, nearly flat relationship likely reflects the diverse cloud substructures and evolutionary states in our sample (e.g., V. Camacho et al. 2020), but highlights a discrepancy between theory and observation in this regime.

A. K. Leroy et al. (2025) and S. E. Meidt et al. (2025) demonstrated that turbulence-regulated star formation models fail to reproduce the observed scaling of star formation efficiency with $\alpha_{\rm vir}$ on GMC scales. Our findings of stable $\alpha_{\rm vir}$ measured across large and small scales argue against the hypothesis that this discrepancy between models and observations stems from scale-dependent bias. These results imply that star formation models may need to consider additional factors besides turbulence and stellar feedback to explain low efficiencies, or that additional terms in the virial equation need to be considered.

#### 4.2.2. Scaling with Mass, Surface Density, and Velocity Dispersion

In Figure 5, we expand our search for signatures of emergent, gravitationally coherent scales by comparing $\alpha_{\rm vir}$ to the mass $M_{\rm CO}$, surface density $\Sigma_{\rm mol}$, and velocity dispersion $\sigma_{\rm v}$ of structures in our catalog.

Plotting $\alpha_{\rm vir}$ against $M_{\rm CO}$ in the left column of Figure 5, we find a weak inverse relationship below $\log_{10}(M_{\rm CO}) \lesssim 7.3$. Excluding the two highest bins, which are dominated by a lineage with little variation in its properties (see the highest-radius bin in Figure 4 and discussion in Section 5.2), and the bin below our completeness limit, a `linmix` fit returns $\alpha_{\rm vir} \propto M_{\rm CO}^{-0.11 \pm 0.03}$. Although $\alpha_{\rm vir}$ scales inversely with $M_{\rm CO}$ in its definition, contributions from structure radii and velocity dispersions clearly mitigate this effect. With the exception of the highest-mass structures, the population of branches appears to continue smoothly from the leaves.

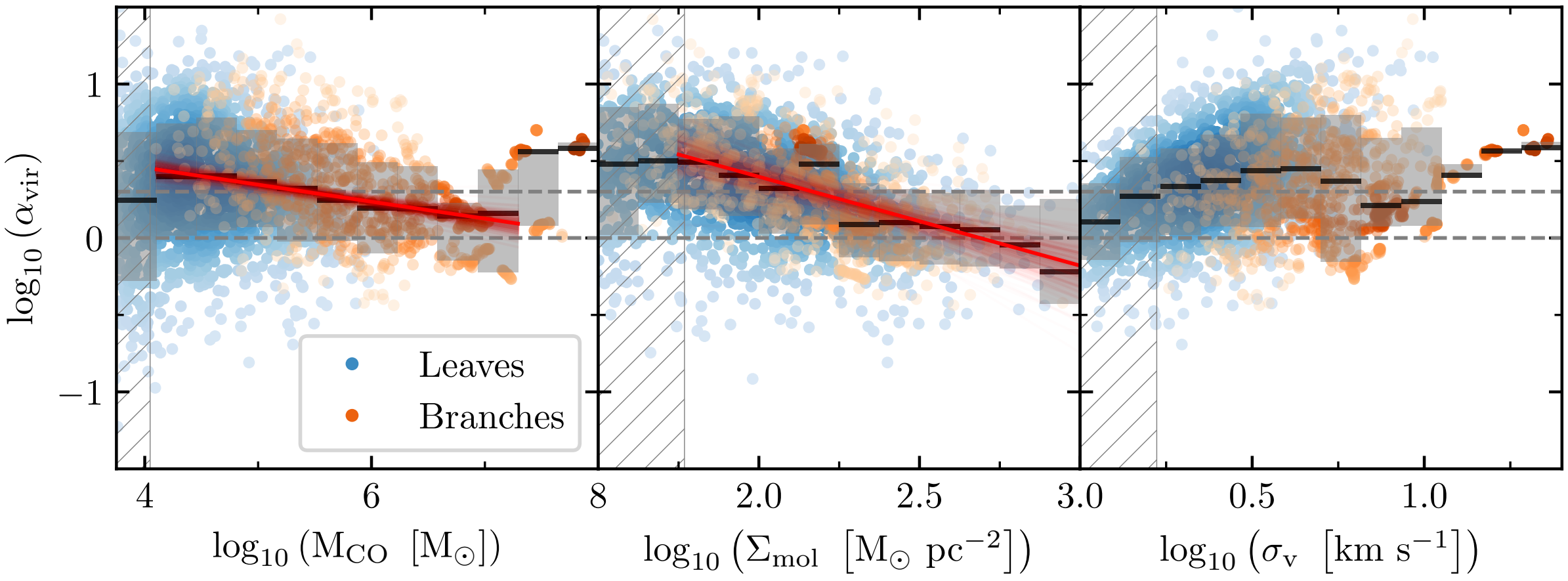

**Figure 5.** Similar to Figure 4 but with $\alpha_{\rm vir}$ plotted against structure mass ($M_{\rm CO}$), surface density ($\Sigma_{\rm mol}$), and velocity dispersion ($\sigma_{\rm v}$) for 12 bins. `linmix` fits above the completeness limits return $\alpha_{\rm vir} \propto M_{\rm CO}^{-0.11 \pm 0.03}$ below $\log_{10}(\rm M_{CO}) \lesssim 7.3$ and $\alpha_{\rm vir} \propto \Sigma_{\rm mol}^{-0.58 \pm 0.16}$.

Numerous relatively massive, unbound structures contribute to flattening our observed $\alpha_{\rm vir}$–$M$ relation. These are similar to those proposed by V. Camacho et al. (2016) and E. Vázquez-Semadeni et al. (2019) as representing early stage cloud collapse, a population also identified in both observations (T. M. Dame et al. 1986; Q. Nguyen-Luong et al. 2016) and simulations (S. A. Mao et al. 2020; R. G. Treß et al. 2021). When compared across consistent mass ranges, several studies show similarly shallow $\alpha_{\rm vir}$–$M$ scaling relations, including J. Roman-Duval et al. (2010); J. Donovan Meyer et al. (2013); D. Colombo et al. (2019) and A. Duarte-Cabral et al. (2021). A.-X. Luo et al. (2024) likewise found a nearly flat relation across six surveys spanning 0.01–100 pc scales, which they attribute to a global balance between gravity and turbulence. Other literature studies report an inverse relationship between $\alpha_{\rm vir}$ and mass, though with varying interpretations. For example, N. J. I. Evans et al. (2021) found higher-mass clouds tend to be more bound across diverse surveys, with many showing $\alpha_{\rm vir} \propto M^{-0.5}$. However, this is the same scaling that M. Chevance et al. (2023) assign to observational censorship from sensitivity and resolution limitations, and tracer-specific biases can lead to an inverse scaling in multitracer analyses (A. Traficante et al. 2018). Our observed, moderately negative relation contrasts with these steeper trends, potentially reflecting the high sensitivity and resolution of our single-tracer data.

Looking to the middle column of Figure 5, we see a clear inverse relationship between $\alpha_{\rm vir}$ and $\Sigma_{\rm mol}$ for intermediate-to-high surface densities, with a slope of $-0.58 \pm 0.16$. The intrinsic scaling of $\alpha_{\rm vir} \propto \Sigma_{\rm mol}^{-1}$ (Equation (6)) contributes significantly. We observe a flattening in this trend below $\log_{10}(\Sigma_{\rm mol}) \lesssim 1.75\, M_\odot\ {\rm pc}^{-2}$, which is intriguing in the context of independent studies suggesting a critical gas surface density of 100–150 $M_\odot\ {\rm pc}^{-2}$ for the onset of star formation (A. Heiderman et al. 2010; C. J. Lada et al. 2010; P. André et al. 2014; N. J. Evans et al. 2014). However, our completeness estimate in Section 3.3 suggests that, in our sample, this turnover is likely attributable to observational bias in the low-completeness regime.

In the right column of Figure 5, we plot $\alpha_{\rm vir}$ against $\sigma_{\rm v}$, finding a mostly flat relation until the highest $\sigma_{\rm v}$ branches. Due to the nature of branches, which contain substructure in physical and/or velocity space by definition, these structures are seen to have higher velocity dispersions.

#### 4.2.3. Scaling with Hierarchical Properties: Height, Level, and Substructure

In Figure 6 we consider the virial parameter's scaling against three quantities intrinsic to the dendrogram structure: the height (peak intensity of a structure excluding its children), level (number of hierarchical steps from the lowest-intensity root, equivalent to the number of direct ancestors), and total number of leaf descendants in a branch's progeny.

The left column of Figure 6, in which we plot $\alpha_{\rm vir}$ against structure height, shows a significant and smooth anticorrelation from the minimum height of 1.0 K to the peak intensity. In other words, more gravitationally bound objects—both leaves and branches—tend to be located at higher intensities, corresponding to regions of brighter CO emission and hence denser gas. Notably, this trend is consistent when plotted for objects in different mass bins, suggesting that it is not primarily driven by the $M^{-1}$ term in $\alpha_{\rm vir}$. These observations are corroborated by the dendrogram visualization (Figure 2) and in panel (a) of our subcontoured distribution (Figure 8).

In the central column of Figure 6, we similarly find a smooth decrease of $\alpha_{\rm vir}$ with structure level, indicating that more bound objects tend to reside in regions of higher hierarchical complexity. This trend is also evident in Figure 2, though degeneracy between the level and height is an important contributor. A high degree of scatter both here and in the $\alpha_{\rm vir}$–height relation suggests that being a bright leaf in a region of complex hierarchy favors, but does not guarantee, boundedness. Correlated chains of structures with stable $\alpha_{\rm vir}$ can be seen to emerge, where branches differ minimally from their parents across several levels due to shared flux (e.g., structures #2118 and #2183 in Figure 2).

Finally, in the right column of Figure 6, we study the relationship between $\alpha_{\rm vir}$ and the total number of leaf descendants of each branch. Below $\log_{10}(\text{desc.}) = 1.9$, there is an intriguing decrease in median $\alpha_{\rm vir}$ with increasing fragmentation. However, above this threshold the trend appears to reverse for the most fragmented structures. Separating the population by radius, we can attribute the turnover to a difference between large and small structures: those with $R < 100$ pc exhibit a significantly negative slope ($m = -0.22 \pm 0.03$), while larger structures (also corresponding to the most fragmented ones) show a positive slope ($m = 0.58 \pm 0.04$). Qualitatively, this suggests that regions of gas with more substructure are more gravitationally bound compared to monolithic regions at similar scales, but only below $R < 100$ pc; above this scale, the correlation between size and degree of fragmentation takes over.

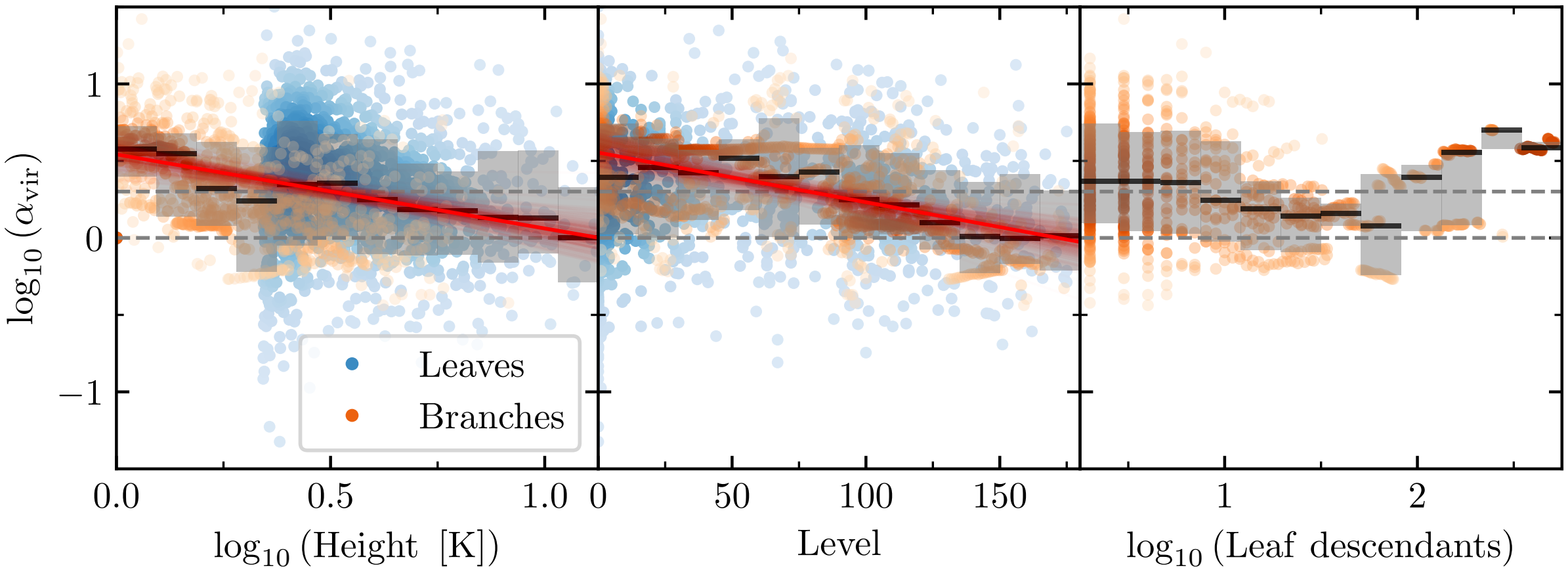

**Figure 6.** Similar to Figures 4 and 5 but with $\alpha_{\rm vir}$ plotted against height (peak value excluding children), hierarchical level, and the total number of leaf progeny. `linmix` fits to all bins return $\alpha_{\rm vir} \propto \text{Height}^{-0.48 \pm 0.10}$ and $\log_{10}(\alpha_{\rm vir}) \propto -(3.2 \pm 0.7)\,\text{Level} \times 10^{-3}$. The feature at $\log_{10}(\text{Height}) = 0.34$ corresponds to the minimum leaf height, i.e., the sum of `min_value` and `min_delta`.

### 4.3. The Virial Parameter Across Hierarchical Mergers

The ratio $R_{\rm vir} = \alpha_{\rm child}/\alpha_{\rm parent}$ parameterizes the change in the virial parameter over structure mergers (Equation (7)), allowing us to study its gradient and scaling relations across the dendrogram hierarchy (see inset panel of Figure 2). When $R_{\rm vir} < 1$, structures are more bound than their parents, while $R_{\rm vir} > 1$ implies the opposite.

In Table 3 and Figure 7's upper right panel, we present the distribution of $R_{\rm vir}$ for leaves and branches. Leaves tend to be slightly more bound than their parents, but both populations are centered around a median value close to $R_{\rm vir} \approx 1$. Branches show a much tighter distribution, which we attribute both to lower uncertainties in the branch population as well as the fact that nested branches can share large fractions of their voxels. The significant scatter we observe explains how $\alpha_{\rm vir}$ can vary dramatically while the median $R_{\rm vir}$ stays constant: although structures are closely related to their parents on average, major changes across individual, scattered mergers tend to lower $\alpha_{\rm vir}$ more significantly than they raise it.

The remaining panels of Figure 7 show $R_{\rm vir}$ plotted against structure parameters as in Figures 4, 5, and 6. With limited exceptions, we see that $R_{\rm vir}$ is mostly flat with a median value close to 1 across all studied quantities. This flat $R_{\rm vir}$ implies largely smooth, linear gradients for the relations of $\alpha_{\rm vir}$ shown in previous figures, when averaged over all lineages. In other words, changes in $\alpha_{\rm vir}$ occur similarly and do not accelerate or decelerate across the scales of our data. A physical (or hierarchical) threshold for the value of $\alpha_{\rm vir}$ would manifest in this space as a cluster of high-magnitude $R_{\rm vir}$ at some characteristic scale, which we do not observe.

**Table 3**
Same as Table 2 but for $R_{\rm vir}$

| Quantity | Leaves | Branches | All |
|---|---|---|---|
| $R_{\rm vir}$ | $0.9^{+1.2}_{-0.5}$ | $1.0^{+0.2}_{-0.2}$ | $1.0^{+0.6}_{-0.5}$ |
| $\lvert\delta R_{\rm vir}/R_{\rm vir}\rvert$ | $1.2^{+0.8}_{-0.5}$ | $0.1^{+0.4}_{-0.1}$ | $0.6^{+1.0}_{-0.6}$ |
| $R_{\rm vir} < 1$ | 56% | 43% | 50% |

For $R$, $M_{\rm CO}$, and $\sigma_{\rm v}$, Figure 7 shows a slightly decreased $R_{\rm vir}$ in low-magnitude bins. Several factors may be contributing to this observed trend: differences between the leaf and branch populations, which are most distinct for these three quantities; the correlation between $R_{\rm vir}$ and $\alpha_{\rm vir}$, which introduces a bias toward low values below our completeness limits; and the substructure of our objects. When a parent branch has three or more components, the decomposition results in two distinct outcomes. First, a single leaf with low $R$ or $\sigma_{\rm v}$, leading to a smaller $\alpha_{\rm vir}$ and $R_{\rm vir}$. Second, a branch containing the remaining substructure, which more closely resembles the parent and therefore has $R_{\rm vir} \approx 1$ (see inset panel of Figure 2). For this reason, very small structures (low $R$, $M_{\rm CO}$, and/or $\sigma_{\rm v}$) tend to have lower $R_{\rm vir}$.

### 4.4. Unextrapolated Scaling within Structures

Our definition of $R_{\rm vir}$ discussed in the previous section considers variations in $\alpha_{\rm vir}$ *across* mergers in the dendrogram, implicitly assuming that these correspond to physically emergent objects. In the following section, we move away from that assumption to explore how dynamical changes can emerge *within* individual dendrogram structures. As the subcontours do not represent full structures, we do not extrapolate quantities to the sensitivity limit of our data.

#### 4.4.1. Internal Scaling with Intensity

Using a modified version of the `levelprops`[28] script, we construct $\Delta I = 0.1$ K contours within each structure from base to peak and recalculate an unextrapolated $\alpha_{\rm vir}$ for all emission contained within each intensity step. In the dendrogram presented in Figure 2, each structure shows these values with a color scale from $0 < \alpha_{\rm vir} < 4$. Visually, this highlights a general decrease in $\alpha_{\rm vir}$ with hierarchical level and with intensity, both between and within individual structures. To explore these trends more explicitly, Figure 8 traces $\alpha_{\rm vir}$ versus intensity level of the contour, where each trace represents the unextrapolated value of $\alpha_{\rm vir}$ as a function of intensity for a single leaf (blue) or branch (orange).

[28] https://gist.github.com/ChrisBeaumont/0c4a1bbca2b748aa41fd

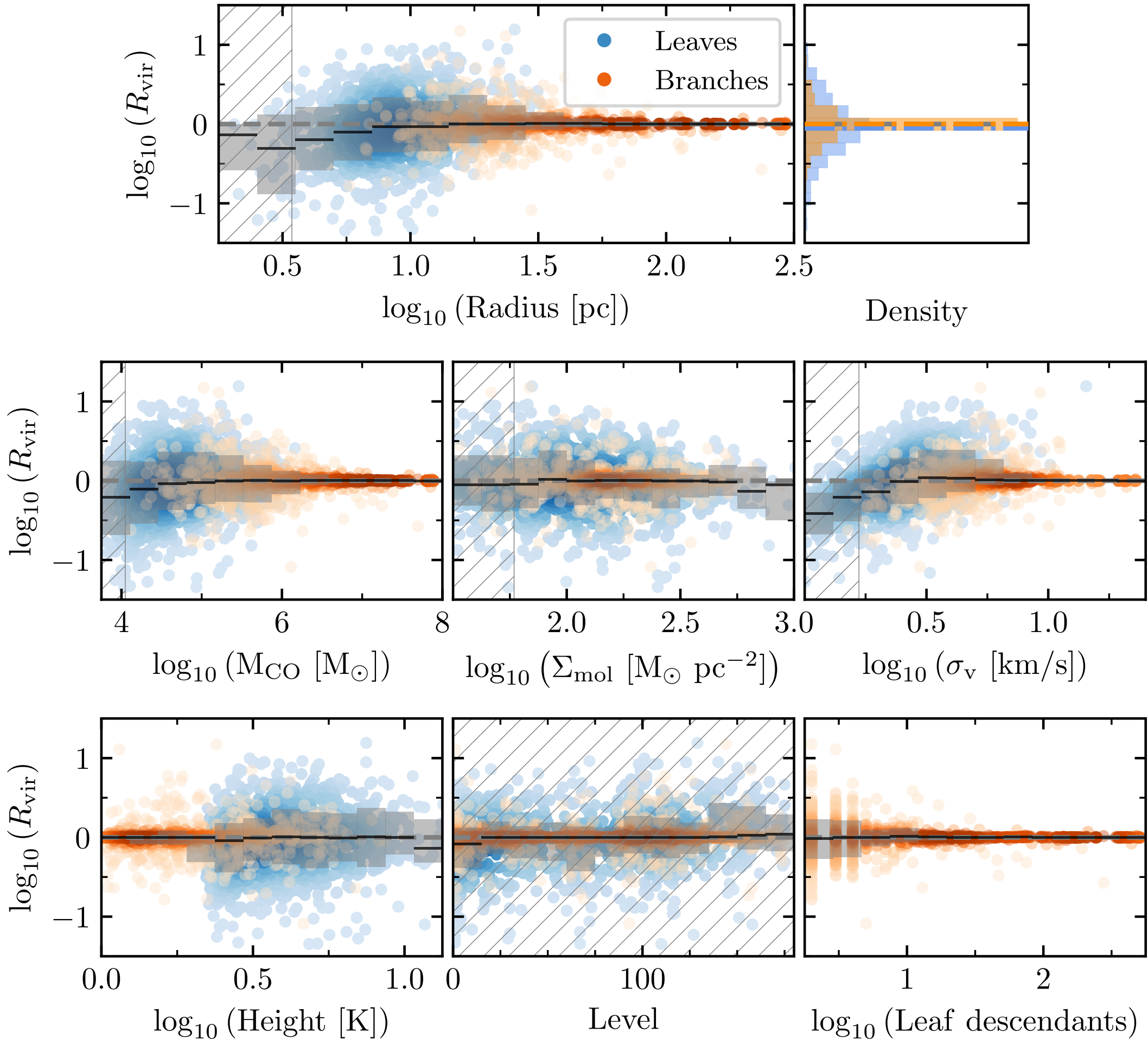

**Figure 7.** The ratio $R_{\rm vir} = \alpha_{\rm child}/\alpha_{\rm parent}$ plotted against structure parameters for all leaves (blue) and branches (orange) in the dendrogram, analogous to Figure 4 (top row), Figure 5 (middle row), and Figure 6 (bottom row). Bins are the same as in those figures and point lightness is similarly scaled by a Gaussian KDE. The dashed, horizontal lines in gray show $\log_{10}(R_{\rm vir}) = 0$, below which structures are more bound than their parents.

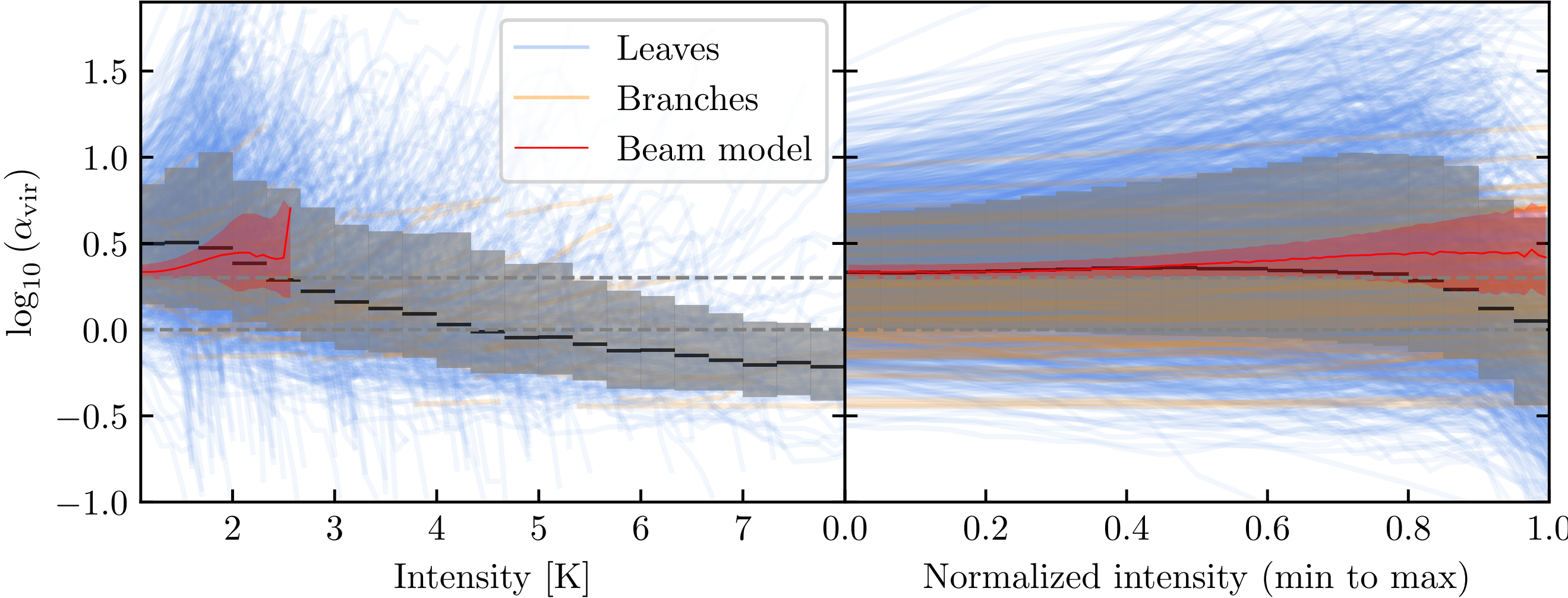

**Figure 8.** Left: unextrapolated virial parameter against intensity for internal traces of individual leaves (blue) and branches (orange). Each trace represents a single structure for which $\alpha_{\rm vir}$ is continuously recalculated as we raise the contour in $\Delta I = 0.1$ K intensity steps. The traces are cut where they contain fewer pixels than half the beam size ($\Omega = 0.2''^2 = 29$ pix). We exclude 55 leaves that span $< 0.5$ K in intensity. Binned medians are shown in black, with their 16–84% ranges in gray. The dashed, horizontal lines show $\alpha_{\rm vir} = 1$ and 2. A beam-sized Gaussian model with realistic noise modeling is shown in red, with the 16–84% range of 2,000 draws shaded. Right: the same as the left panel, but normalized from the minimum to the maximum intensity of each structure.

To compare these traces with the observed beam structure, we follow M. H. Heyer et al. (2001) to construct a model described by the following Gaussian distribution along the angular and spectroscopic coordinates:

$$T(x, y, v) = T_0 \exp(-2.77((x/R_x)^2 + (y/R_y)^2 + (v/\delta v)^2)) + T_N. \quad (9)$$

We set $R_x$ and $R_y$ equal to the FWHM of our beam, estimate $\delta_v$ from a subsequent fit to the discrete size–linewidth relation (Figure 10), and draw $T_N$ from a distribution of Gaussian noise with variance $\sigma^2$. At $5.5\sigma$, we choose the amplitude $T_0$ to be just above our detection limit while ensuring $\alpha_{\mathrm{vir},0} \approx 2$, the median of the normalized trace sample. The median levelprops result for 2000 draws of this model cloud is shown in Figure 8.

The left panel of Figure 8 shows a relatively steady decrease in median $\alpha_{\mathrm{vir}}$ as we increase in absolute intensity. When we normalize across the full extent of our structures in the right panel, we see that the changes in $\alpha_{\mathrm{vir}}$ tend to be smooth (or small) and hence flat for most of this space. The normalized sample median is steady and consistent with our beam model at $\alpha_{\mathrm{vir}} \approx 2$ up until ~80% of the structure maximum. Most structures seem to follow a similar shape, if not magnitude, to the beam model below this threshold.

Both of these relations are subject to a large degree of scatter across structures that is primarily due to variations in the initial value of $\alpha_{\mathrm{vir}}$. A second major contribution to the observed scatter is dramatic increases or decreases ("spikes") in the value of $\alpha_{\mathrm{vir}}$ near the peak intensities. For the increases, our intensity-weighted second moment approach to calculating structure radii and velocity dispersions may overestimate $R$ and $\sigma_{\mathrm{v}}$ for structures with multiple spatial or spectral components, raising $\alpha_{\mathrm{vir}}$. We confirmed this by modifying Equation (3) to use a simple geometric definition for structure radius, $R' = \sqrt{A/\pi}$. Using this definition significantly suppresses the increasing spikes. Decreasing spikes can be explained by stochasticity from low number statistics at the highest contour levels, which leads to some very bright pixels that drive $\alpha_{\mathrm{vir}}$ downwards near structure peaks. Experimenting with stricter volume limits than $\Omega/2$ significantly reduces the number of decreasing spikes. We visually inspected individual spiking structures at different intensity levels to validate both of these interpretations.

Overall, Figure 8 shows a significant decrease in $\alpha_{\mathrm{vir}}$ as we move to higher intensities. When normalized across the intensity span, however, $\alpha_{\mathrm{vir}}$ appears steady and its variation relatively smooth for most structures, with the notable "spiking" exceptions explained by mathematical or statistical effects. Indeed, only about half (56%) of leaves have lower values of $\alpha_{\mathrm{vir}}$ at their highest-intensity contour than at their lowest, and very few (12%) decrease monotonically across their full intensity extent. Taken together with the structure-based analyses presented in Figures 6 and 7, these results imply that major changes in gravitational boundedness over short intensity spans typically come from structures merging, rather than from changes within individual structures.

The subcontouring approach we take in Figure 8 is comparable to the "differential virial analysis" adopted by M. R. Krumholz et al. (2025) and C. J. Lada et al. (2025), who show that looking within clouds can probe divergent predictions for cloud collapse. For M31's GMCs, they found that virial parameters increased toward regions of high surface density in a majority (64%) of cloud-scale structures. This characteristic increase, they argue, points to structures that are supported against collapse by internal turbulence rather than in a state of global dynamical collapse. N. Peretto et al. (2023) similarly found increasing $\alpha_{\mathrm{vir}}$ toward infrared-dark cloud centers on scales of $\sim$ 0.3–30 pc, although their structure profiles appear to flatten as they approach $R \gtrsim 10$ pc. Our results, which show a comparable increase in just 44% of structures, are in slight tension with both these observations. This discrepancy may be attributable to differences in the observed size scales, which emphasizes the need for additional high-resolution observations with multiscale decompositions that incorporate subcontouring approaches.

#### *4.4.2. The Size–Linewidth Relation within Structures*

A scaling relation between molecular cloud sizes and linewidths of the form $\sigma_{\mathrm{v}} \propto R^{\beta}$ was identified by R. B. Larson (1981) and has since been foundational for observational studies of turbulence in the molecular ISM (M. Heyer & T. M. Dame 2015). M. H. Heyer & C. M. Brunt (2004) linked this relation to the first-order velocity structure function, which encodes the turbulent scaling driving spatial and velocity fluctuations in the observed gas. They identified a near-constant scaling within and between 27 GMCs of different sizes, environments, and star formation activities, pointing toward universal, compressible turbulence in molecular clouds.

Applying the levelprops subcontouring to our data in Figure 9, we extend a similar study of turbulent scaling over a much larger sample. This figure shows the unextrapolated size–linewidth relation with traces for each leaf and branch as in Figure 8. Each lowest-level contour is shown as a point, equivalent to the unextrapolated $R$ and $\sigma_{\mathrm{v}}$ of the dendrogram structure in question.

Following M. H. Heyer & C. M. Brunt (2004), we apply least-squares fits to these data using three different approaches: first, we fit the unextrapolated lowest-contour points as a population; second, we include all subcontoured traces, treating them as a unified population; and third, we fit each structure's individual trace, weighted by the number of voxels per contour. The resulting fits are shown in Figure 9. Their remarkable consistency supports an invariant, universal turbulence law, as described by M. H. Heyer & C. M. Brunt (2004), when averaged over our large population. However, the scatter in Figure 9 and large standard deviation over the individual traces suggest significant variation between structures at a level much higher than they propose, mediated by the significant spread in surface densities apparent in Figure 11.

### *4.5. Extrapolated Dynamical Scaling Relations*

#### *4.5.1. The Size–Linewidth Relation*

In Figure 10, we show the extrapolated size–linewidth relation for leaves (blue) and branches (orange) in our data. The observed size–linewidth scaling reflects the interplay between surface density, virial parameter, and the turbulent properties of structures (Equations (5) and (6); M.-M. Mac Low & R. S. Klessen 2004; C. F. McKee & E. C. Ostriker 2007; J. Ballesteros-Paredes et al. 2011). For a population with approximately constant $\alpha_{\mathrm{vir}}$, the size–linewidth scaling directly connects to surface density variations through the "Heyer–Keto" relation discussed in Section 4.5.2. Like R. Shetty et al. (2012), we find uniform scaling from sizes of a few to several hundred parsecs, indicating a fundamental connection between dense structures and their surrounding gas. Cascading compressible turbulence (A. G. Kritsuk et al. 2013; R. Cen 2021) and gravitational influences (C. L. Dobbs et al. 2011; J. C. Ibáñez-Mejía et al. 2016) have been invoked to steepen the canonical relation, consistent with what we observe. Critically, our result leverages the full hierarchical sample's dynamic range; looking at leaves alone produces a substantially weaker correlation, agreeing with R. Shen et al. (2024) who similarly found robust scaling relations when incorporating hierarchical structures.

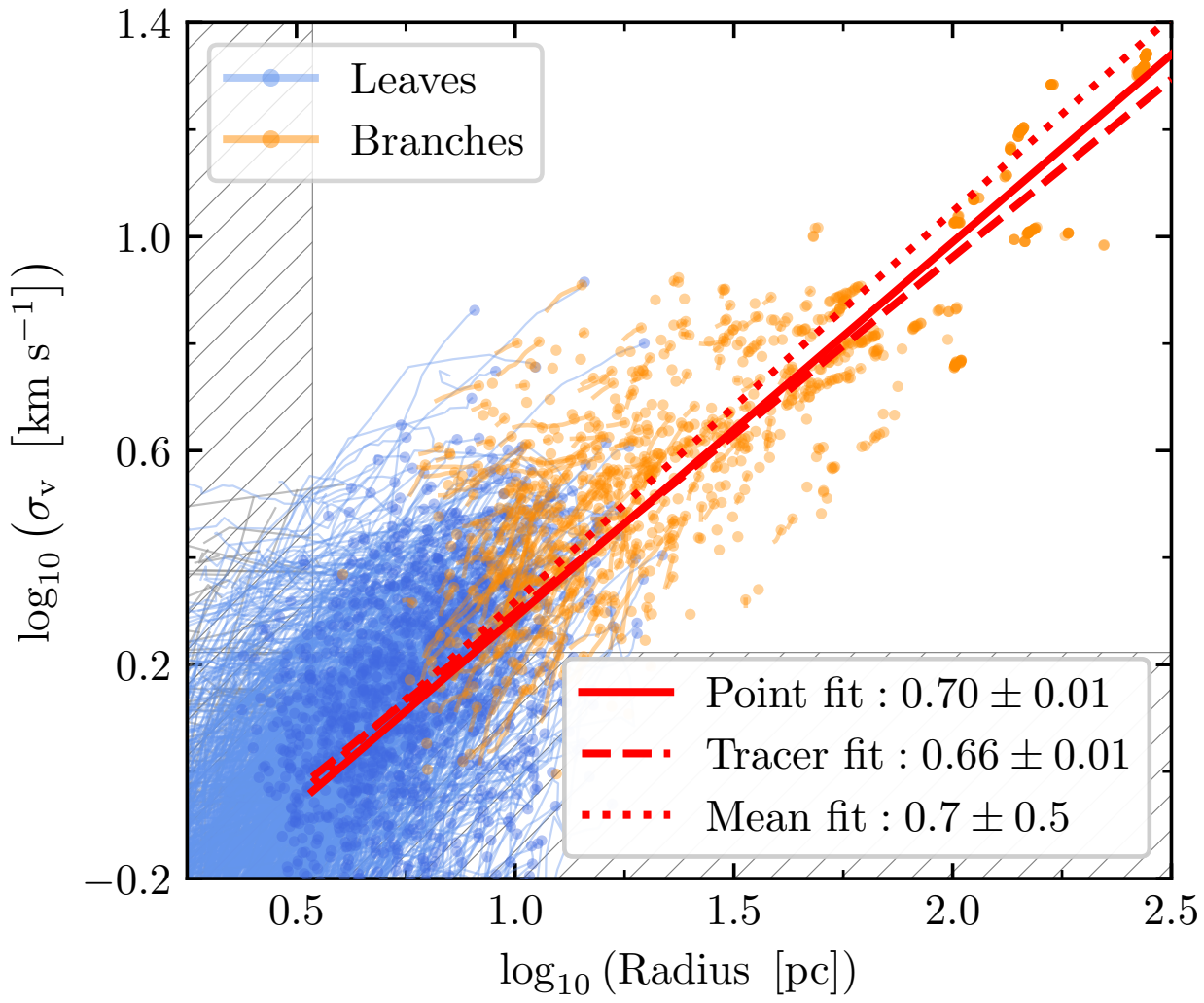

**Figure 9.** Structure velocity dispersion against radius for all leaves (blue) and branches (orange), but this time for unextrapolated quantities (points). We additionally use `levelprops` traces as in Figure 8 to show the change of velocity dispersion and size as the contour level is raised in 0.1 K intensity steps. The traces are plotted in gray where they contain fewer pixels than half the beam size. Three linear least-squares fits are shown, with their slopes in km $s^{-1}$ $pc^{-1}$: to the unextrapolated points (solid line; uncertainty is standard error), all points including traces (dashed line; uncertainty is standard error), and the mean of fits to individual traces (dotted line; uncertainty is standard deviation). The estimated 50% completeness limits (Section 3.3) are demarcated by the hatched boundaries.

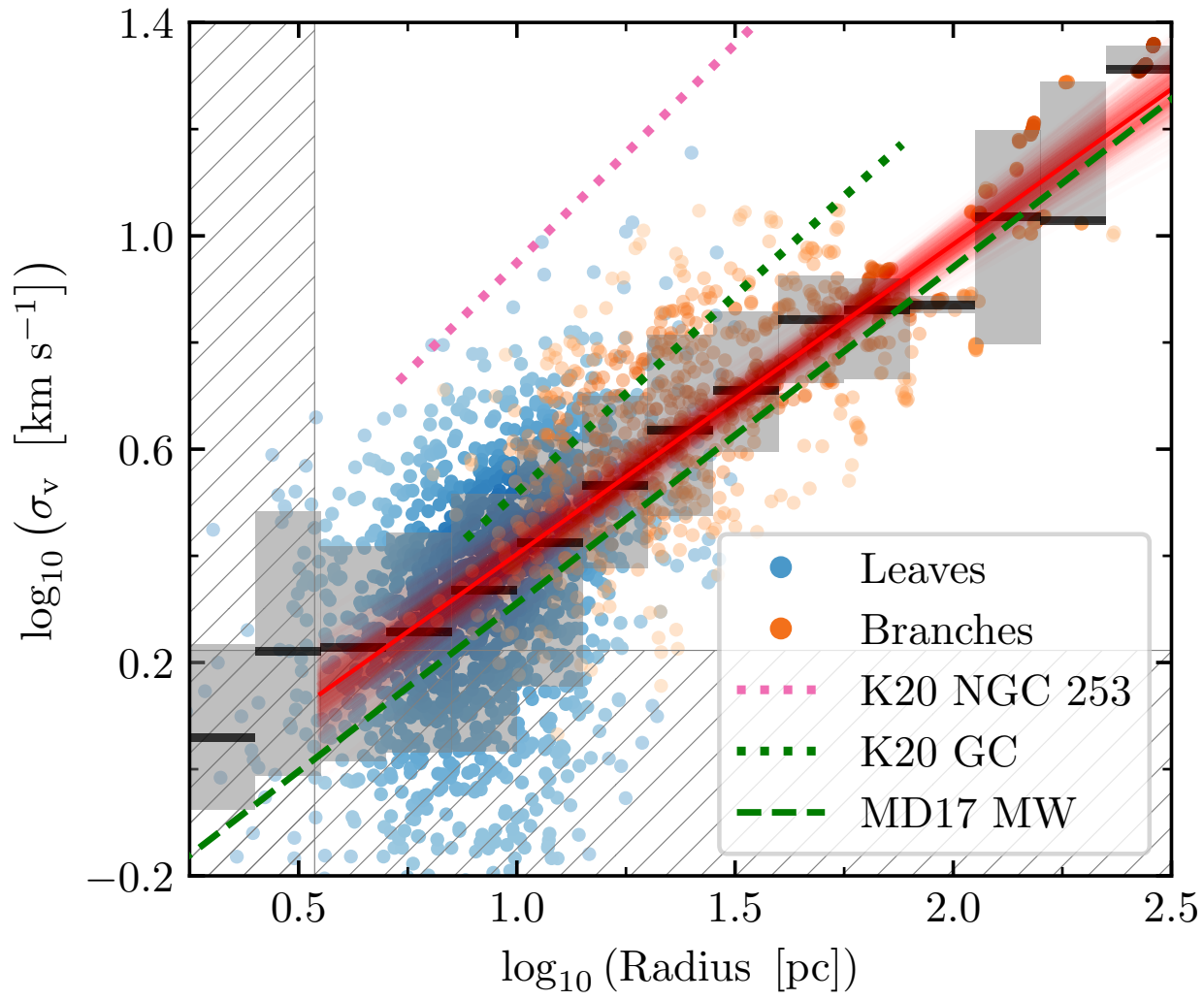

**Figure 10.** Structure velocity dispersion against radius for all leaves (blue) and branches (orange) in the dendrogram. Each point's lightness is scaled by a KDE as in Figure 4. Bins of 0.15 dex are shown with their medians in black and 16–84% ranges in gray. A `linmix` fit to the centers of bins above the completeness limit gives $\sigma_v \propto R^{0.58 \pm 0.04}$, plotted in red. We show N. Krieger et al.'s (2020) CO(1 − 0) fitted relations for the center of NGC 253 (pink) and the Milky Way (green) as dotted lines, with M.-A. Miville-Deschênes et al.'s (2017) fit for the Milky Way disk as a green dashed line. The estimated 50% completeness limits (Section 3.3) are demarcated by the hatched boundaries.

Our measured slope of 0.58 ± 0.04 is slightly steeper than, though within $2\sigma$ of, the canonical coefficient $\beta \approx 0.5$ established for Milky Way clouds (e.g., T. M. Dame et al. 1986; P. M. Solomon et al. 1987; C. F. McKee & E. C. Ostriker 2007; P. García et al. 2014; T. S. Rice et al. 2016). Despite these earlier works establishing $\beta \approx 0.5$, our result joins several recent studies suggesting $\beta$ may be closer to 0.6 or higher (e.g., M.-A. Miville-Deschênes et al. 2017; M. Benedettini et al. 2020; L. Liu et al. 2021; J.-X. Zhou et al. 2022; T. Colman et al. 2024; Y. Sun et al. 2024). Y.-H. Xie & G.-X. Li (2025) argue that this steepening reflects the dual influence of gravity and galactic shear with an approximate transition regime of ∼100 pc. While the steeper slope is consistent with these results, we do not see a smooth transition in moving from small to large scales in our data that would indicate an obvious break from turbulence- to shear-dominated regimes.

To compare our results in Figure 10 to environments with distinct gas conditions, we additionally plot the relations in $^{12}$CO(1–0) from N. Krieger et al. (2020) for the center of NGC 253 and the galactic center as well as M.-A. Miville-Deschênes et al.'s (2017) relation for the Milky Way disk. Given the slope uncertainties (as high as 0.30 in the values quoted by M.-A. Miville-Deschênes et al. 2017) and the large scatter between structures in each data set, most of the variation we observe seems to come from differences in structure velocity dispersions at a given size, which can also be influenced by resolution and tracer-dependent effects, rather than systematic changes to the slope (N. Krieger et al. 2020). The consistency of our findings with M.-A. Miville-Deschênes et al.'s (2017) sample, despite our extragalactic context, suggests fundamental similarities in how turbulence scales across different environments. Elevated linewidths in the central molecular zones of both NGC 253 and the Milky Way are reflected in the corresponding literature relations.

#### *4.5.2. The Heyer–Keto Relation*

For a known virial parameter, Equations (5) and (6) imply a simple scaling between the quantity $\sigma_v/R^{0.5}$ and $\Sigma_{\rm mol}$ (the "Heyer–Keto" relation, E. R. Keto & P. C. Myers 1986; M. Heyer et al. 2009; M. Heyer & T. M. Dame 2015). We plot these parameters for leaves (blue) and branches (orange) in the left panel of Figure 11. Using the approximation

$$P_{\rm turb} = \rho\sigma_v^2 = \frac{3\Sigma_{\rm mol}\sigma_v^2}{4R}, \tag{10}$$

we trace the isobars for several internal pressures, from $P_e/k_B = 10^2\ {\rm K\ cm^{-3}}$ to $10^8\ {\rm K\ cm^{-3}}$ (G. B. Field et al. 2011; J. Sun et al. 2018).

In the right panel of Figure 11, we additionally compare the density distribution of our leaves to three catalogs: E. Rosolowsky et al. (2021), who identified 4986 GMCs across 10 galaxies of the $^{12}$CO(2–1) PHANGS-ALMA survey at 90 pc resolution, M. Heyer et al. (2009), who reanalyzed the Galactic molecular clouds of P. M. Solomon et al. (1987) in $^{13}$CO(1–0) at $\leqslant$3 pc resolution, and J. Sun et al. (2020), who analyzed

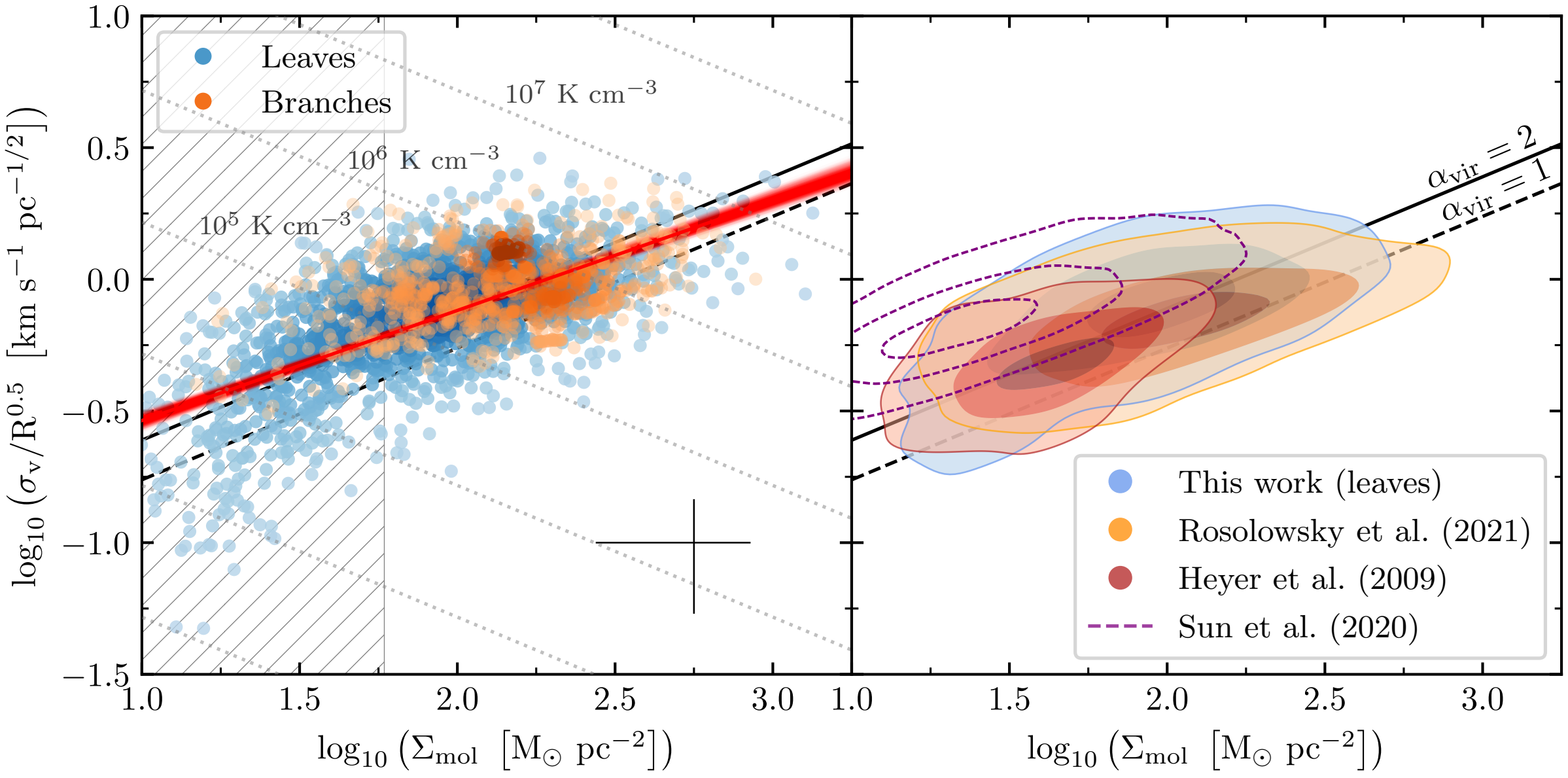

**Figure 11.** Left: Heyer–Keto plot showing the variation of $\sigma_v/R^{0.5}$ with surface density $\Sigma_{\rm mol}$. Leaves are plotted in blue and branches in orange, with their brightness scaled by a Gaussian KDE reflecting the smoothed density distribution. Dotted gridlines indicate fixed $P_{\rm turb}$ with 1 dex spacing. The dashed and solid black lines represent the boundaries for structures in virial equilibrium ($\alpha_{\rm vir}=1$) and energy equipartition ($\alpha_{\rm vir}=2$) in the absence of external pressure. A `linmix` regression to the leaf sample, giving $\sigma_v/R^{0.5} \propto \Sigma_{\rm mol}^{0.42\pm0.01}$, is shown in red. The median bootstrapped leaf uncertainty is given by an error bar in the bottom right. Right: comparing the Heyer–Keto relation in our leaf sample against the literature catalogs of E. Rosolowsky et al. (2021; 90 pc resolution, extragalactic GMCs with $^{12}$CO), M. Heyer et al. (2009; $\leqslant 3$ pc resolution, Galactic GMCs with $^{13}$CO), and J. Sun et al. (2020; 90 pc, extragalactic beam-scale measurements with $^{12}$CO). Each distribution is smoothed and visualized with a 2D Gaussian KDE at the 16%, 50%, and 84% density contours. Structural decompositions, unlike the beam-scale approach, occupy a similar region of the parameter space despite observing at very different scales.

independent, 90 pc-scale sight lines in 35 PHANGS-ALMA galaxies.

Our leaves exhibit a very similar distribution compared to the extragalactic GMC catalog of E. Rosolowsky et al. (2021) and scale consistently with the Milky Way clouds of M. Heyer et al. (2009), despite significant differences in resolution, characteristic structure scale, and galactic environment. This consistency across heterogeneous data sets with a wide range of physical resolutions, from $\leqslant$3 pc to 90 pc, is noteworthy and recalls our finding of approximately constant $\alpha_{\rm vir}$ across all measured scales in Figure 4. On the other hand, J. Sun et al.'s (2020) beam-scale, structure-agnostic decomposition produces lower $\Sigma_{\rm mol}$ and higher $\sigma_v/\mathrm{R}^{0.5}$, which could reflect the inclusion of lower-density gas that is not associated with individual gas structures, as well as beam smearing of unresolved velocity gradients (T. G. Williams et al. 2023).

The power-law slope we observe is lower than that expected for clouds with a fixed dynamical state and self-gravity in approximate balance with kinetic energy, i.e., $\sigma_v/R^{0.5} \propto \Sigma_{\rm mol}^{0.5}$ (E. Schinnerer & A. K. Leroy 2024). This is closely related to and reflects the correlation $\alpha_{\rm vir} \propto \Sigma_{\rm mol}^{-0.58}$ seen in Figure 5. The dominance of external pressure over self-gravity would manifest as a flattening or inversion of the $\sigma/R^{0.5}$–$\Sigma_{\rm mol}$ relation at low $\Sigma_{\rm mol}$ (G. B. Field et al. 2011; J. Sun et al. 2018), which we do not observe. Similar populations of molecular gas structures with elevated linewidths have been observed in galaxy centers (e.g., T. Oka et al. 2001; J. Sun et al. 2018; J. Sun et al. 2020; N. Krieger et al. 2020; C. Battersby et al. 2025) as well as in regions of low gas surface density, where they are sometimes referred to as "chaff" (M. H. Heyer et al. 2001; V. Camacho et al. 2016). These may be attributed variously to observational biases, external pressure, unresolved streaming motions, transience, or the effects of galactic potentials (S. E. Meidt et al. 2018). The absence of such low-surface-density, high-velocity-dispersion structures in Figure 11 may be due to the environment we target, with $20 \lesssim \Sigma_{\rm mol} \lesssim 600\,M_\odot\,\mathrm{pc}^{-2}$ and more than a kiloparsec removed from the starbursting galaxy center or the low-density outer galaxy (our Figure 1; also see K. Sakamoto et al. 2011; A. K. Leroy et al. 2015, 2018). The relatively high sensitivity ($\sim 5\,M_\odot\,\mathrm{pc}^{-2}$) and velocity resolution ($\sim 1\,\mathrm{km\,s^{-1}}$) of these data (Table 1) also enable us to resolve structures that might otherwise appear blended and artificially broadened in lower-sensitivity surveys.

## 5. Discussion

### *5.1. A Synthetic Picture of Hierarchical Boundedness*

Our analyses of the dendrogram catalog in Section 4 suggest a population of marginally bound structures with extrapolated $\alpha_{\rm vir} \approx 2$ across all size scales. Leaves, which are free of substructure and sit at the top of the dendrogram hierarchy, tend to be marginally more bound than branches (Table 2). Notably, we do not find evidence that the gravitational boundedness of molecular gas changes with spatial scale over the range we study; that is, median $\alpha_{\rm vir}$ is approximately constant from $R \sim 3$ to 250 pc (Figure 4).

More gravitationally bound structures are more massive, denser, and have lower velocity dispersions than less bound structures (Figure 5). We observe a significant population of massive, unbound structures that themselves contain significant substructure, which may represent early stage cloud collapse. Being a bright leaf in a region of complex hierarchy favors, but does not guarantee, boundedness—more bound structures are also higher in their peak intensity and position in the dendrogram hierarchy, but $\alpha_{\rm vir}$ is not clearly related to the amount of substructure observed in a given region (Figure 6).

Leaves tend to be slightly more bound than their parent structures, but child branches do not change substantially (parameterized by $R_{\rm vir}$; Figure 7). We do not observe increases in the rate of change of $\alpha_{\rm vir}$ as a function of scale or other physical characteristic; $R_{\rm vir}$ does not significantly correlate with any parameter we study except $\sigma_{\rm v}$, which we explain as an effect of the decomposition algorithm. Although the median change in $\alpha_{\rm vir}$ relative to the hierarchy is small ($\langle R_{\rm vir}\rangle \approx 1$), the large scatter suggests that individual objects can still differ greatly from their parents, and that this difference is almost as likely to lead to more boundedness as less.

$\alpha_{\rm vir}$ and $R_{\rm vir}$ exhibit a high degree of scatter at every size scale. However, we see mostly smooth, flat changes when we look within individual structures as a function of intensity (Figure 8). This suggests that structure-to-structure mergers within the dendrogram account for most of the variation in $\alpha_{\rm vir}$. Less than half of our leaves increase in $\alpha_{\rm vir}$ with intensity, in slight tension with the results of M. R. Krumholz et al. (2025).

Accepting that $\alpha_{\rm vir}$ is an approximate, if imperfect, tracer of the dynamical state of molecular gas, our data undermine a view of bound clumps within larger, unbound structures, or at least suggest that dynamical signatures of such clumps are only observable at $R < 3$ pc (e.g., M. Benedettini et al. 2020; T. Colman et al. 2024; K. Wang et al. 2024). The stability of the median $\alpha_{\rm vir}$ between small- and large-scale structures is consistent with the global hierarchical collapse model, in which signatures of gravitational collapse would be expected on all scales (E. Vázquez-Semadeni et al. 2024b). It could also reflect a highly heterogeneous population of structures—or a continuous, hierarchical ISM—where a number of proposed processes take place in different regions and across different scales.

### 5.2. Structure on Large Scales

In the rightmost bins of Figures 4 and 5, several lineages of exceptionally large size ($R \sim$ 200–300 pc), mass ($M_{\rm CO} \sim 10^{7.5-8.0}\,M_\odot$), and velocity dispersion ($\sigma_{\rm v} \sim$ 15–20 km s$^{-1}$) are apparent, which result in a relatively high $\alpha_{\rm vir} \approx 4$. While not universal (some structures in this regime have a more typical $\alpha_{\rm vir}$), the apparent increase suggests a possible transition regime of $R \gtrsim 100$ pc in which $\alpha_{\rm vir}$ begins to increase significantly. Given the large size scales involved, these structures encompass regions of relatively faint emission such that they have a fairly typical $\Sigma_{\rm mol}$ and location in the size–linewidth relation (Figure 10).

To extend the regime probed to the largest scales available in our data, we estimate the characteristic virial parameter for our entire field by treating the cube as a single structure. We apply the strict mask described in Section 2.3 and calculate $\alpha_{\rm vir}$ as in Section 3.2, finding $\alpha_{\rm vir} = 8.5$ for an equivalent size scale of $R = 400$ pc, significantly higher than typical (smaller) structures in our catalog.

For objects on large scales ($R \gtrsim 100$ pc), gas kinematics may become dominated by galactic orbital motions (e.g., M. Krumholz & A. Burkert 2010; A. Hughes et al. 2013; S. M. R. Jeffreson et al. 2021; Y.-H. Xie & G.-X. Li 2025), and stellar mass can significantly influence the gravitational potential (S. E. Meidt 2022). Streaming motions associated with the galactic potential can also affect the balance of gas pressure, increasing the stability of the largest structures within spiral arms (S. E. Meidt et al. 2013, 2018, 2020, 2025) and contributing meaningfully to observed velocity dispersions (Y.-H. Xie & G.-X. Li 2025). Therefore, it is highly likely that the largest structures in our catalog are significantly affected by galactic shear and that our estimate of $\alpha_{\rm vir}$ includes systemic motion over the larger region while neglecting important contributions to the gravitational potential.

In Figure 4, this dynamics-driven increase in $\alpha_{\rm vir}$ does not appear to exhibit a single, universal transition scale. This suggests that, while shear and galactic rotation have important effects on measurements of $\sigma_{\rm v}$ and $\alpha_{\rm vir}$, there may not be a universal transition scale (for example, at the galaxy's scale height) but rather one that depends on the geometry and internal structure of the molecular gas and may vary smoothly as structure sizes increase.

### 5.3. Limitations of the Observer's Virial Parameter

Our bootstrapped uncertainties on $\alpha_{\rm vir}$ are described in Section 4.1, and we briefly review possible systematics here.

The projected nature of our observations introduces intrinsic scatter in $\sigma_{\rm v}$ and $R$ (C. N. Beaumont et al. 2013; M. Y. Grudić et al. 2019; S. Cahlon et al. 2024), while opacity broadening can contribute to increased $\sigma_{\rm v}$ (A. Hacar et al. 2016). Our construction of $\alpha_{\rm vir}$, which follows the standard formulation, does not consider gravitational surface terms (J.-G. Kim et al. 2021), tidal forces (L. Ramírez-Galeano et al. 2022; J. Lee et al. 2025), an external gravitational potential (S. E. Meidt et al. 2018; L. Liu et al. 2021), pressure confinement (M. H. Heyer et al. 2001), or magnetic support (J.-G. Kim et al. 2021). As discussed in the previous section, high values of $\alpha_{\rm vir}$ for large structures are better interpreted as a combination of gravito-turbulent support and galactic shear, rather than accurately diagnosing the level of boundedness of the gas.

More sophisticated formulations of $\alpha_{\rm vir}$ than Equation (6) have been proposed by these authors and others, but our focus on scaling relationships and interoperability with the literature makes the standard approach well-suited for this analysis. Our results depend on the systematic variation of $\alpha_{\rm vir}$ with scale rather than its absolute magnitude across the structure population, which makes our conclusions robust to many of these effects. While interpreting the exact value of $\alpha_{\rm vir}$ as a precise indicator of the dynamical state of individual gas structures requires careful consideration of systematic uncertainties, studying how $\alpha_{\rm vir}$ varies with size or density scale, as we do in this work, mitigates such challenges (M. R. Krumholz et al. 2025).

## 6. Summary and Conclusions

We have presented a high-resolution ALMA survey combining 12 m, 7 m, and TP observations survey in $^{12}$CO(2–1) over a 1.4 kpc × 5.6 kpc field in the star-forming disk of NGC 253. These $0''.4$-resolution data provide a fully sampled view of the molecular ISM from scales of 7 pc to more than a kiloparsec (Figure 1). Our final data cube has a sensitivity of 0.89 K km s$^{-1}$ or 5.3 $M_\odot$ pc$^{-2}$ (Table 1).

We use a dendrogram algorithm to identify multiscale, hierarchical structure in the data. This yields a catalog of 963 branches and 1500 leaves, presented in Figure 2. For each structure in this catalog, we derive the radius, velocity dispersion, mass, surface density, and virial parameter, following moment methods and correcting for sensitivity and resolution bias via extrapolation and deconvolution. We estimate a 50% completeness limit of $1.1 \times 10^4\ M_\odot$ (Figure 3).

Our primary goal is to search for a characteristic scale of gravitational boundedness, thereby testing the hypothesis of an ISM structured around discrete, bound clouds. Our analysis finds that:

1. The median $\alpha_{\rm vir}$ and 16%–84% range is $2.0^{+2.8}_{-1.2}$ for leaves and $2.2^{+2.0}_{-1.1}$ for branches. 48% of our leaves and 46% of our branches appear gravitationally bound, with $\alpha_{\rm vir} < 2$ (Table 2).
2. The median $\alpha_{\rm vir}$ is largely independent of structure radius from $R \sim 3$ to 200 pc (Figure 4). Above this scale, $\alpha_{\rm vir}$ gradually increases as galactic dynamics play a greater role, but we do not observe a distinct transition threshold.
3. Gravitationally bound structures tend to be more massive, denser, and have lower velocity dispersions (Figure 5). While $\alpha_{\rm vir}$ decreases for structures that are high in their peak intensity and position in the hierarchy, it does not depend on the amount of substructure contained within branches (Figure 6).
4. The population-averaged scaling of $\alpha_{\rm vir}$ with physical and hierarchical quantities is smooth, showing no inflection points or discontinuities above our estimated completeness limits (Figure 7).

Looking within individual structures, we introduce the `levelprops` approach to estimate their properties as a function of CO intensity. This analysis reveals:

1. $\alpha_{\rm vir}$ decreases with increasing intensity, but most of this effect occurs near structure peaks and across hierarchical mergers (Figure 8).
2. The turbulent size–linewidth scaling is invariant with an exponent of $\beta \approx 0.7$ when averaged over our population, but shows large scatter between individual structures (Figure 9).

We also examine extrapolated, multiscale dynamical relations and compare these to literature results, finding:

1. The size–linewidth (Figure 10) and Heyer–Keto (Figure 11) relations hold across our hierarchical sample, and reflect consistent properties for literature structure catalogs across substantially different resolutions and environments.
2. We find a relatively steep size–linewidth coefficient of $\beta = 0.58 \pm 0.04$ and a shallow Heyer–Keto scaling of $0.42 \pm 0.01$.

Ultimately, the absence of emergent, gravitationally bound scales in our data suggests that nonhierarchical cloud decompositions may impose arbitrary sizes on the ISM and supports a view of the ISM as a continuous, hierarchical medium. Our exceptional data bridge the gap between cloud-scale extragalactic surveys and studies of clumps on parsec scales, but additional high-resolution observations of molecular gas, even pushing to parsec scales, are needed to determine whether these findings hold at even finer physical resolution and across a broader range of galactic environments.

**Acknowledgments**

This work was carried out as part of the PHANGS collaboration.

We thank the anonymous referee for a thoughtful and detailed review. Special thanks to J. Ballesteros-Paredes, C. Beaumont, J. den Brok, E. Ostriker, and E. Vázquez-Semadeni for insightful comments and discussions.

This research makes use of astrodendro, a Python package to compute dendrograms of astronomical data (http://www.dendrograms.org/)

This paper makes use of the following ALMA data, which have been processed as part of the PHANGS-ALMA CO(2–1) survey:

ADS/JAO.ALMA#2017.1.01101.S.

ADS/JAO.ALMA#2018.1.00596.S.

ALMA is a partnership of ESO (representing its member states), NSF (USA), and NINS (Japan), together with NRC (Canada), MOST and ASIAA (Taiwan), and KASI (Republic of Korea), in cooperation with the Republic of Chile. The Joint ALMA Observatory is operated by ESO, AUI/NRAO and NAOJ. The National Radio Astronomy Observatory is a facility of the National Science Foundation operated under cooperative agreement by Associated Universities, Inc.

E.R. acknowledges the support of the Natural Sciences and Engineering Research Council of Canada (NSERC), funding reference number RGPIN-2022-03499. S.C.O.G. acknowledges financial support from the European Research Council via the ERC Synergy Grant "ECOGAL" (project ID 855130), from the German Excellence Strategy via the Heidelberg Cluster of Excellence (EXC 2181 - 390900948) "STRUCTURES." J.S. acknowledges support by the National Aeronautics and Space Administration (NASA) through the NASA Hubble Fellowship grant HST-HF2-51544 awarded by the Space Telescope Science Institute (STScI), which is operated by the Association of Universities for Research in Astronomy, Inc., under contract NAS 5-26555.

*Facilities:* ALMA, No:45m.

*Software*: Astropy (Astropy Collaboration et al. 2013 2018, 2022), linmix (B. C. Kelly 2007), Matplotlib (J. D. Hunter 2007), NumPy (C. R. Harris et al. 2020), SciPy (P. Virtanen et al. 2020), spectral-cube (A. Ginsburg et al. 2015).

## Appendix A
## ALMA Observation Log

Table 4 presents detailed information regarding observations for the high-resolution ALMA field that represents the main data set used in this work. In the table, we report the project code, array and configuration (when relevant), the number of antennas used, the date and duration of the observation, the average elevation, the precipitable water vapor (PWV), and the minimum and maximum baseline length (for interferometric observations) associated with the data. Individual execution blocks typically lasted ∼40–90 minutes, with a mean elevation at 54–83°and PWV between 0.4 and 1.8 mm, reflecting near-optimal conditions with the exception of one TP execution block. Our 7 m baselines span ∼9–49 m while the 12 m baselines span ∼15–1398 m.

**Table 4**
ALMA Observation Log for Each Execution Block

| Project | Array | $N_{ant}$ | Start (MJD) | End (MJD) | $t_{on-source}$ (min) | Avg. elev. (deg) | PWV (mm) | *u-v* range (m) |
|---|---|---|---|---|---|---|---|---|
| 2018.1.00596.S | 12 m/C43-5 | 50 | 58421.1714 | 58421.1722 | 40.82 | 77.3 | 0.89 | 15.1–1397.8 |
| 2018.1.00596.S | 12 m/C43-5 | 47 | 58422.1849 | 58422.1856 | 40.82 | 72.1 | 1.02 | 15.1–1397.8 |
| 2018.1.00596.S | 12 m/C43-5 | 46 | 58425.0475 | 58425.0482 | 40.82 | 65.2 | 1.78 | 15.1–1397.8 |
| 2018.1.00596.S | 12 m/C43-2 | 43 | 58563.6473 | 58563.6478 | 27.22 | 59.4 | 0.90 | 15.0–313.7 |

**Note.** The full table, showing TP and 7 m observations, is available as a machine-readable table. Columns include Project–the ALMA project code; Array–the array and configuration of the observation; $N_{ant}$–the number of antennas participating in the observation; Start and End–the beginning and end of the observations reported as Modified Julian Day (00:00 on 2018 January 1 is MJD 58119); Avg. elev.–the mean elevation during the observation; PWV–the precipitable water vapor during the observation; *u-v* range–the minimum and maximum baseline.

(This table is available in its entirety in machine-readable form in the online article.)

## Appendix B
## The CO-to-$H_2$ Conversion Factor and $R_{21}$

The CO-to-$H_2$ conversion factor $\alpha_{CO}$ has been shown to vary with environment and the local conditions of molecular gas, particularly the metallicity, velocity dispersion, density, and temperature (e.g., M. G. Wolfire et al. 2010; R. Shetty et al. 2011; R. Feldmann et al. 2012; D. Narayanan et al. 2012; M. Gong et al. 2020). This variation can be attributed to changes in the fraction of CO-dark molecular gas as well as the emissivity and opacity of CO-emitting regions (E. Schinnerer & A. K. Leroy 2024). We therefore consider several prescriptions for $\alpha_{CO}$: namely, the standard Milky Way factor of $\alpha_{CO,MW}^{(1-0)} = 4.35\ M_\odot\ (\mathrm{pc^2\ K\ km\ s^{-1}})^{-1}$ (A. D. Bolatto et al. 2013), approaches that incorporate the velocity dispersion or metallicity, and a dust-based calibration in NGC 253.

To do so, we derive relevant quantities from the inclination-corrected "mega-tables" catalog of J. Sun et al. (2022) and J. Sun et al. (2023), averaging values within the three 1.5 kpc hexagonal bins that overlap with our field. For a metallicity-dependent conversion factor, we define

$$\alpha_{CO}^{(1-0)} = \alpha_{CO,MW}^{(1-0)} \cdot (Z/Z_\odot)^{-1.5}, \tag{B1}$$

where a scaling exponent of 1.5 is a broadly adopted value that agrees with calibrations using dust and depletion times (e.g., G. Accurso et al. 2017; L. K. Hunt et al. 2020; E. Schinnerer & A. K. Leroy 2024). With $\langle Z \rangle = 1.12\ Z_\odot$, this gives $\alpha_{CO}^{(1-0)} = 3.68\ M_\odot\ (\mathrm{pc^2\ K\ km\ s^{-1}})^{-1}$. I.-D. Chiang et al. (2024) used the dust surface density inferred from far infrared emission and a constant dust-to-metals ratio to calibrate $\alpha_{CO}$'s correlation with the stellar mass surface density. Their Equation 19, with $\langle \Sigma_\star \rangle = 372\ M_\odot\ \mathrm{pc^{-2}}$, implies $\alpha_{CO}^{(1-0)} = 2.76\ M_\odot\ (\mathrm{pc^2\ K\ km\ s^{-1}})^{-1}$ when combined with the metallicity correction above. I.-D. Chiang et al. (2024) also directly report intensity-weighted mean values for $\alpha_{CO}$. For a more detailed view, we use a resolved catalog of measurements from I. Chiang (2025, private communication) sampled over a $0\farcm6$ grid, whose mean value for the overlapping regions is $\alpha_{CO}^{(1-0)} = 2.30\ M_\odot\ (\mathrm{pc^2\ K\ km\ s^{-1}})^{-1}$. Finally, Y.-H. Teng et al. (2024) propose a correction for the cloud-scale velocity dispersion, which also gives $\alpha_{CO}^{(1-0)} = 2.30\ M_\odot\ (\mathrm{pc^2\ K\ km\ s^{-1}})^{-1}$ using $\langle \langle \Delta v \rangle_{150\mathrm{pc}} \rangle = 7.1\ \mathrm{km\ s^{-1}}$.

Ultimately, we adopt a constant $\alpha_{CO}^{(1-0)} = 3.68\ M_\odot\ (\mathrm{pc^2\ K\ km\ s^{-1}})^{-1}$ (Equation (B1)) to account for a mildly super-solar metallicity expected for the region and allow for ease of comparison to other works. We choose to forgo a starburst correction based on the stellar mass surface density or cloud-scale velocity dispersion due to their derivation on larger scales and difficulties in independently tracing the total gas mass.

The CO(2–1)-to-CO(1–0) line ratio $R_{21}$ is also known to vary with the local ISM conditions, particularly $\Sigma_{SFR}$ (J. S. den Brok et al. 2021; A. K. Leroy et al. 2022; J. den Brok et al. 2025). Accordingly, E. Schinnerer & A. K. Leroy (2024) recommend

$$R_{21}(\Sigma_{SFR}) = 0.65 \times \left(\frac{\Sigma_{SFR}}{1.8 \times 10^{-2}\ M_\odot\ \mathrm{yr^{-1}\ kpc^{-2}}}\right)^{0.125}, \tag{B2}$$

which gives $R_{21} = 0.600$ for $\langle \Sigma_{SFR} \rangle = 9.37 \times 10^{-3}\ M_\odot\ \mathrm{yr^{-1}\ kpc^{-2}}$ as calculated above.

We additionally measure $R_{21}$ directly using $^{12}$CO(1 − 0) data from the Nobeyama 45 m telescope (K. Sorai et al. 2000; N. Kuno et al. 2007) in conjunction with $^{12}$CO(2 − 1) ACA observations previously mentioned (A. K. Leroy et al. 2021b). To do so, we smooth, convolve, and resample the Nobeyama and ACA cubes to a common grid, following procedures outlined in E. Rosolowsky & A. Leroy (2006) to generate signal masks and calculate the velocity-integrated intensities. The median value of the cubes' ratio over our field returns $R_{21} = 0.615$, consistent with the $\pm 0.18$ scatter of the E. Schinnerer & A. K. Leroy (2024) calibration. This same value is recovered when we mask the moment 0 maps to the approximate galactocentric annulus of the field, 100″–300″. The stability of our $R_{21}$ estimate is encouraging, and we use this value going forward. Taking $\alpha_{CO}^{(1-0)} = 3.68\ M_\odot\ (\mathrm{pc^2\ K\ km\ s^{-1}})^{-1}$ estimated above, we therefore adopt $\alpha_{CO}^{(2-1)} = 5.99\ M_\odot\ (\mathrm{pc^2\ K\ km\ s^{-1}})^{-1}$ as the best value for our observations (Equation (4)).

We conservatively adopt half the range in $\alpha_{CO}^{(1-0)}$ values, $0.69\ M_\odot\ (\mathrm{pc^2 K\ km\ s^{-1}})^{-1}$, as the uncertainty on this quantity. For $R_{21}$, we take the $\pm 0.18$ scatter quoted by E. Schinnerer & A. K. Leroy (2024). With these values, the uncertainty in our estimated $\alpha_{CO}^{(2-1)}$ is 35%, dominated by the $R_{21}$ term.

## Appendix C
## The Impact of Dendrogram Parameter Choices

We study the effect of our parameter choices on the dendrogram and our science results by varying each while holding the other two constant, using:

(i) `min_value` $= \{1.5, 2, 2.5, 3\ 3.5\} \cdot \sigma$, `min_delta` $= 3 \cdot \sigma$, `min_npix` $= 4 \cdot \Omega$;

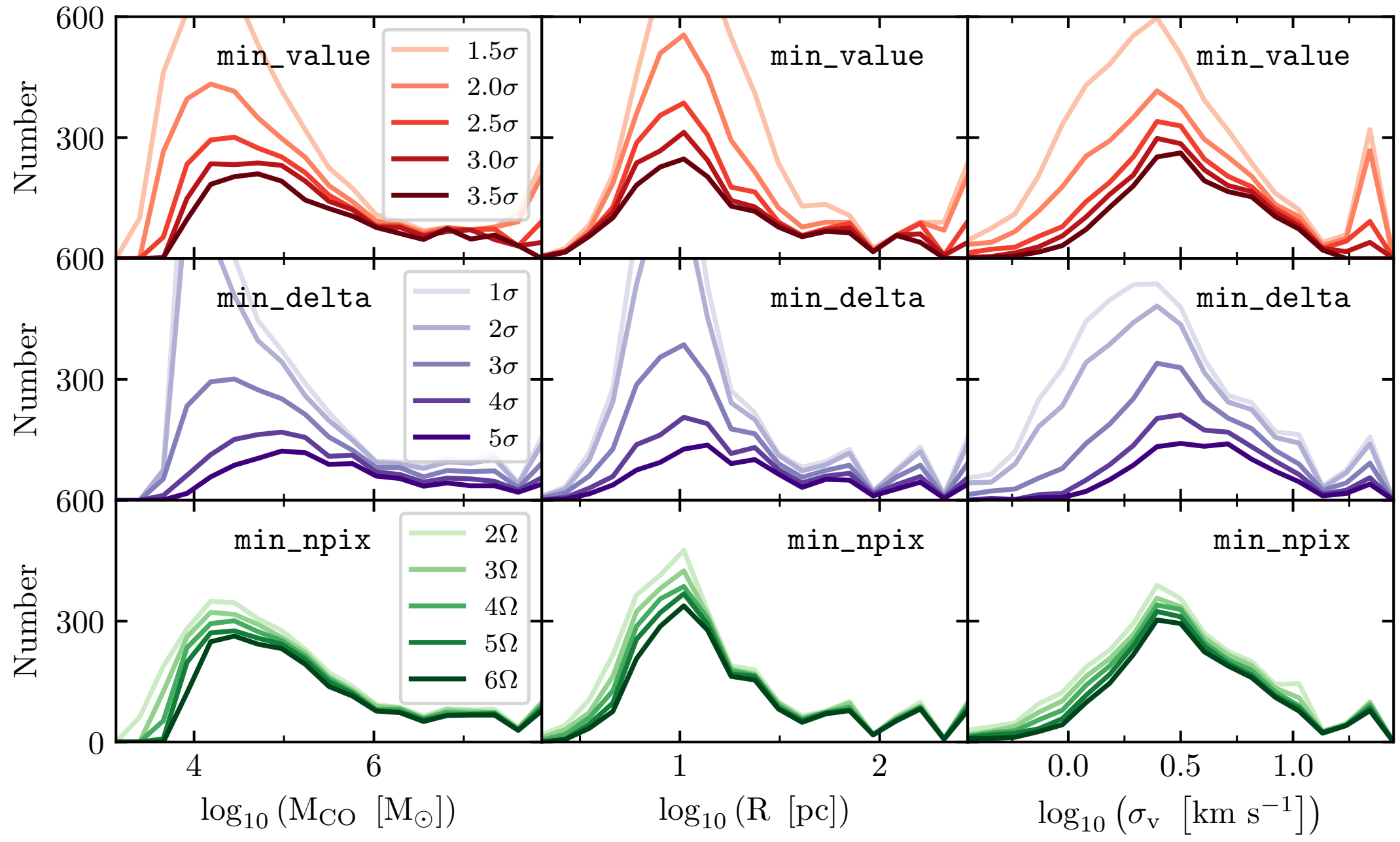

**Figure 12.** Exploring the impact of dendrogram parameter choices on the resulting mass, radius, and velocity dispersion distribution functions. The canonical parameters are set to `min_value` $= 2.5\sigma$, `min_delta` $= 3\sigma$, and `min_npix` $= 4\Omega$, where $\sigma$ represents the characteristic noise of 0.4 K and $\Omega$ the beam solid angle in pixels (see Section 3.1). In the top row (red) we vary `min_value` while keeping the other two parameters constant; in the middle (purple) we vary `min_delta`; and in the bottom (green) we vary `min_npix`.

(ii) `min_delta` $= \{1, 2, 3, 4, 5\} \cdot \sigma$,
`min_value` $= 2.5 \cdot \sigma$, `min_npix` $= 4 \cdot \Omega$;
(iii) `min_npix` $= \{2, 3, 4, 5, 6\} \cdot \Omega$, `min_value` $= 2.5 \cdot \sigma$,
`min_delta` $= 3 \cdot \sigma$,

where $\sigma$ represents the characteristic noise of 0.4 K and $\Omega$ the beam solid angle in pixels (Section 3.1).

Varying each of the parameters in the range that we studied significantly affects the resulting hierarchy and number of structures in the dendrogram. In Figure 12, we compare the effects of these choices on the mass, radius, and velocity dispersion distribution functions of the structure catalog. In addition to these basic properties, we seek to understand the impact of our parameter choices on science results in Figure 13. We focus on the slope of Bayesian `linmix` fits to three of our key scaling relations, binned within consistent ranges: the virial parameter with radius (Figure 4, Section 4.2.1), size–linewidth (Figure 10, Section 4.5.1), and Heyer–Keto (Figure 11, Section 4.5.2).

Increasing `min_value` raises the minimum intensity floor, which causes low-intensity leaves and branches to drop out of the dendrogram (Figure 12). For the range that we studied, the number of dendrogram structures changes from 5370 to 1450 and the fraction of leaf structures remains nearly constant, at 61% $\pm$ 1%. By breaking up the largest branches and removing small, low-velocity-dispersion structures, we see in Figure 13 that the $\alpha_{\rm vir}$–$R$ slope decreases abruptly. Our $\sigma_{\rm v}$–$R$ relation is flattened and the Heyer–Keto slope becomes slightly less significant due to the loss of a low-surface-density, low-velocity-dispersion population.

Increasing `min_delta` raises the minimum significance between structures, so that small peaks will be added to their neighboring structures rather than identified as discrete objects. High values also excise low-intensity leaves so that for small structures the effects mirror high values of `min_value`. For the range that we studied, the number of structures changes from 4856 to 1011 and the fraction of leaves ranges from 66% to 57%, reflecting a decrease in branching and loss of isolated leaves. The slope of the $\alpha_{\rm vir}$–$R$ relation remains fairly constant, as the small leaves that are eliminated tend to scatter evenly about the median value of $\alpha_{\rm vir}$. The $\sigma_{\rm v}$–$R$ and Heyer–Keto slopes are qualitatively affected similarly as for `min_value` due to a loss of low-surface-density and low-radius, low-velocity-dispersion structures.

Within the range that we studied, `min_npix` appears to have a much smaller effect on the dendrogram and our analysis. High values excise the smallest structures but leave the larger hierarchy relatively stable, slightly increasing the median mass, radius, and velocity dispersion. The number of structures decreases from 2974 to 2062, and the leaf fraction shrinks marginally, from 62% to 61%. The slopes for each of the three fits remain consistent within the statistical errors.

For low values of each parameter, we note a small spike in the number of very high-mass, -size, and -velocity-dispersion structures in Figure 12. This reflects the expansion of the two largest roots, #488 and #87 (Figures 1 and 2, green and cyan, respectively) into the low-intensity regions around them, which increases the number of very large branches.

This analysis shows that reasonable choices of `min_value` and `min_delta` can significantly affect the physical properties of a dendrogram catalog, while `min_npix` has far less power. These effects are most important for low-mass, low-radius, and low-velocity-dispersion structures. Choices of `min_value` below $2.5\sigma$ can significantly increase scatter, particularly in the $\alpha_{\rm vir}$–$R$ relation, while higher values of `min_value` and `min_delta` introduce limitations due to lower completeness. Taking the maximum standard deviation across each panel in Figure 13, the resulting biases introduce a systematic error to the $\alpha_{\rm vir}$–$R$, $\sigma_{\rm v}$–$R$, and Heyer–Keto slopes of 0.12, 0.09, and 0.11, respectively. This is an important consideration as we compare our values to the literature, but does not affect the conclusions we draw in this paper.

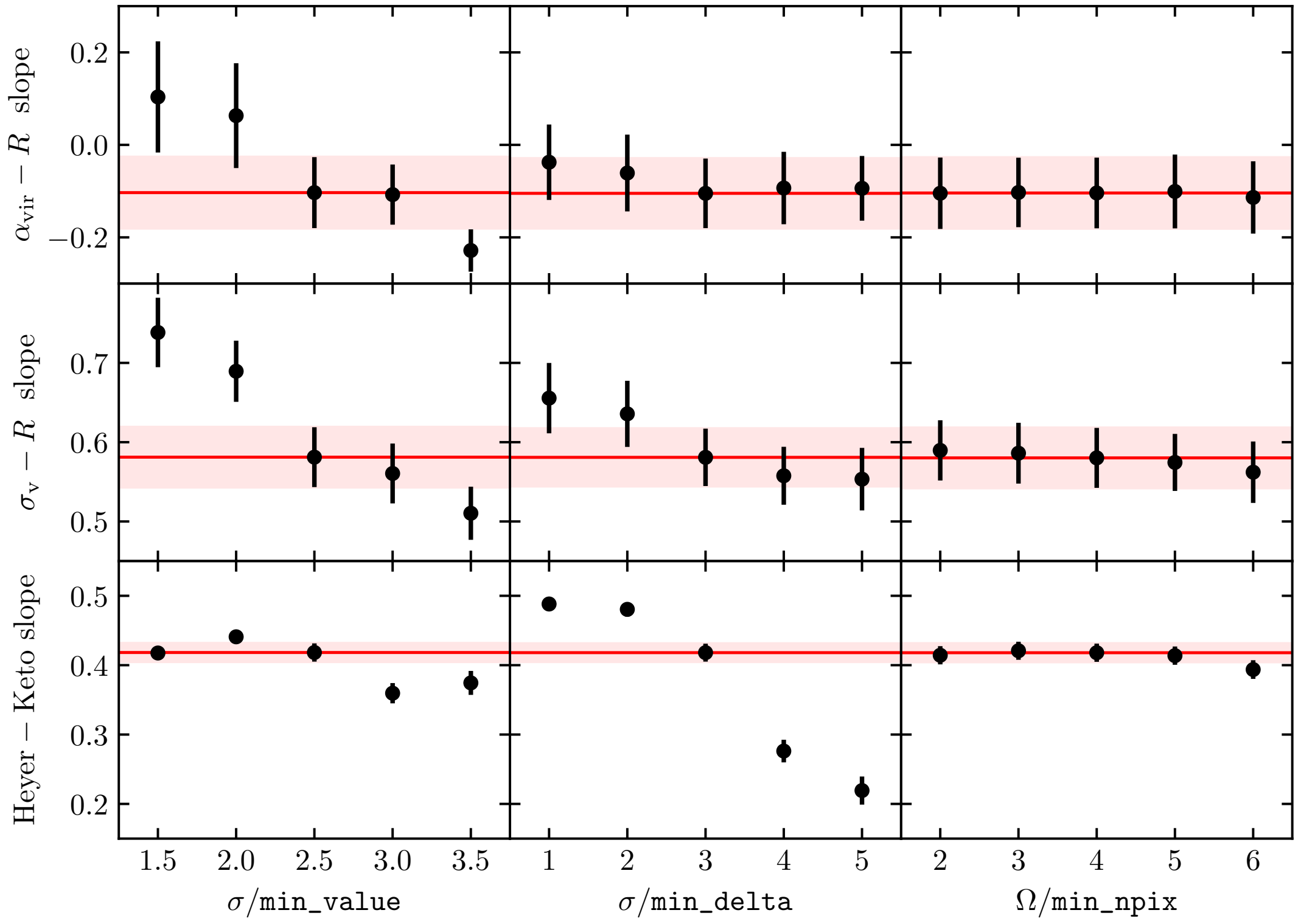

**Figure 13.** Linear regression `linmix` fit slopes resulting from varying dendrogram parameters. Each row shows results for a different scaling relation: $\alpha_{\rm vir}$–$R$ (Figure 4), $\sigma_{\rm v}$–$R$ (Figure 10), and Heyer–Keto (Figure 11). For each column, we have varied the parameter on the $x$-axis while keeping the others constant. Canonical parameter values are the same as in Figure 12, with the corresponding slopes and standard deviations marked by the red horizontal lines and shaded regions.

## Appendix D
## Aspect Ratios and Elongated Structure

As an additional parameter to characterize the geometries of our objects, we define the aspect ratio as the ratio between the intensity-weighted second spatial moment along the major and minor axes, $\sigma_{\rm maj}/\sigma_{\rm min}$. As a population, leaves (median = 1.9) have a slightly lower aspect ratio than branches (median = 2.2), reflecting their generally more compact, symmetrical shapes in position–position space. In Figure 14, we plot aspect ratio against size for leaves and branches, finding no statistically significant correlation. This suggests that the more rounded nature of leaves is not simply driven by their relatively compact sizes, but is closely related to the lack of substructure within individual objects. Applying a dendrogram decomposition to 3D dust extinction maps in the Milky Way, Y.-H. Xie et al. (2024) found that larger structures tend to be more elongated. Apparent discrepancies between these results could be due to different systematics in the data (e.g., elongation effects due to distance uncertainties) or the fact that each approach is tracing different components of the ISM. Additionally, Y.-H. Xie et al. (2024) define their canonical cloud size along the longest axis, while we use a geometric mean (our Equation (3)) that naturally leads to a less significant correlation.

On the other hand, aspect ratios derived from our moment approach (Equation (2)) may poorly describe the morphology of some complex, filamentary structures (about 15% of our branches, by eye). Other approaches such as medial axis skeletonization (e.g., K. R. Neralwar et al. 2022) are promising for these large structures, but their characterization as a single filament or cloud is questionable and a more detailed analysis of their morphologies is beyond the scope of this work.

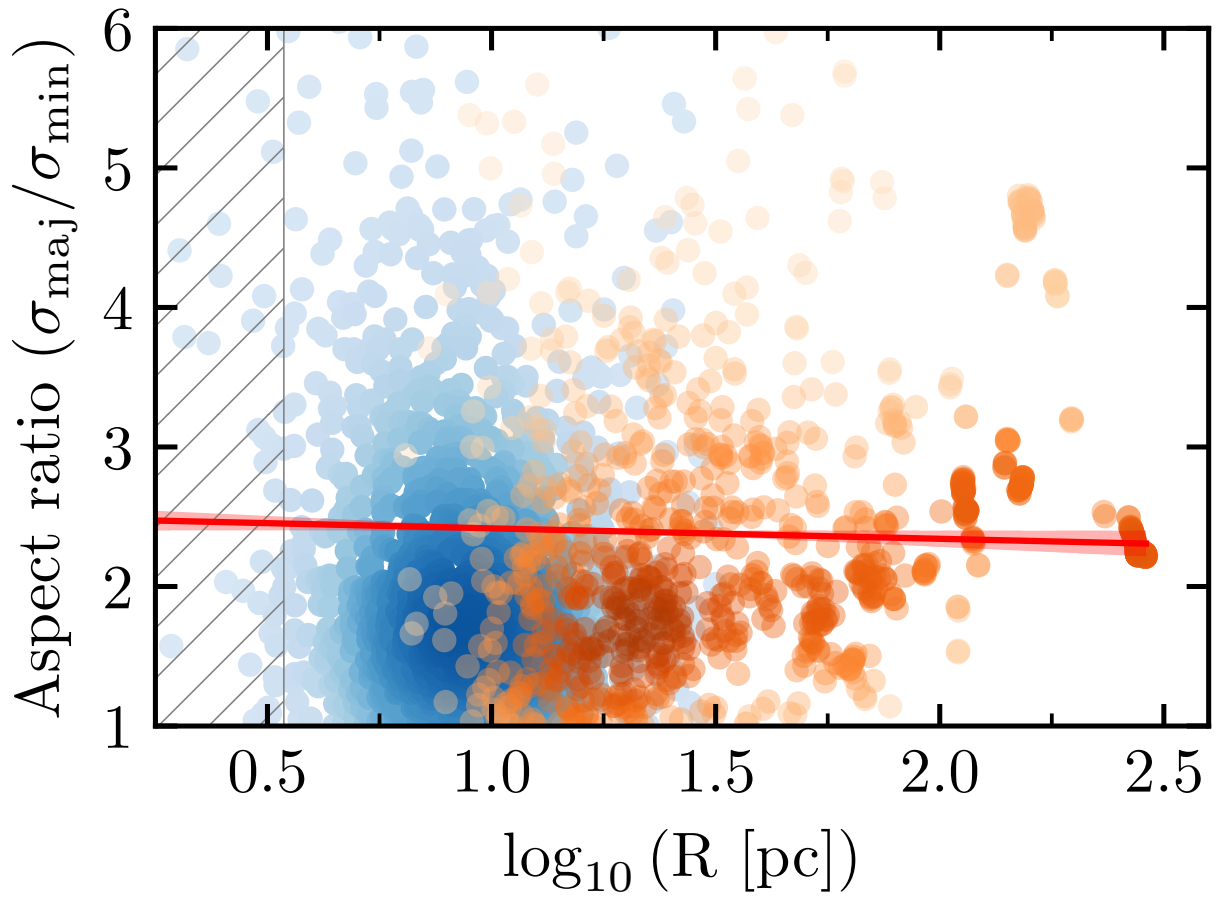

**Figure 14.** Structure aspect ratio against the logarithm of size for leaves (blue) and branches (orange) in our catalog. A least-squares fit, shown in red, returns a slope of $-0.1 \pm 0.1$ pc$^{-1}$.

In Figure 15, we plot the logarithm of $\alpha_{\rm vir}$ against radius as in Figure 4, but colored by the aspect ratio. Evidently, the aspect ratio of structures is not a significant contributor to the scaling or scatter of the virial parameter with other quantities presented in Figures 4, 5, and 6, which reflects our choice of radius and velocity dispersion diagnostics as per F. Bertoldi & C. F. McKee (1992). We also do not observe any statistically significant correlations (least-squares fits are consistent with 0) between the aspect ratio of structures and their mass, surface density, velocity dispersion, height, or level.

Galactic shear, in concert with gravitational confinement of molecular gas in the disk, can lead to structures elongated along the direction of Keplerian rotation (e.g., G.-X. Li et al.

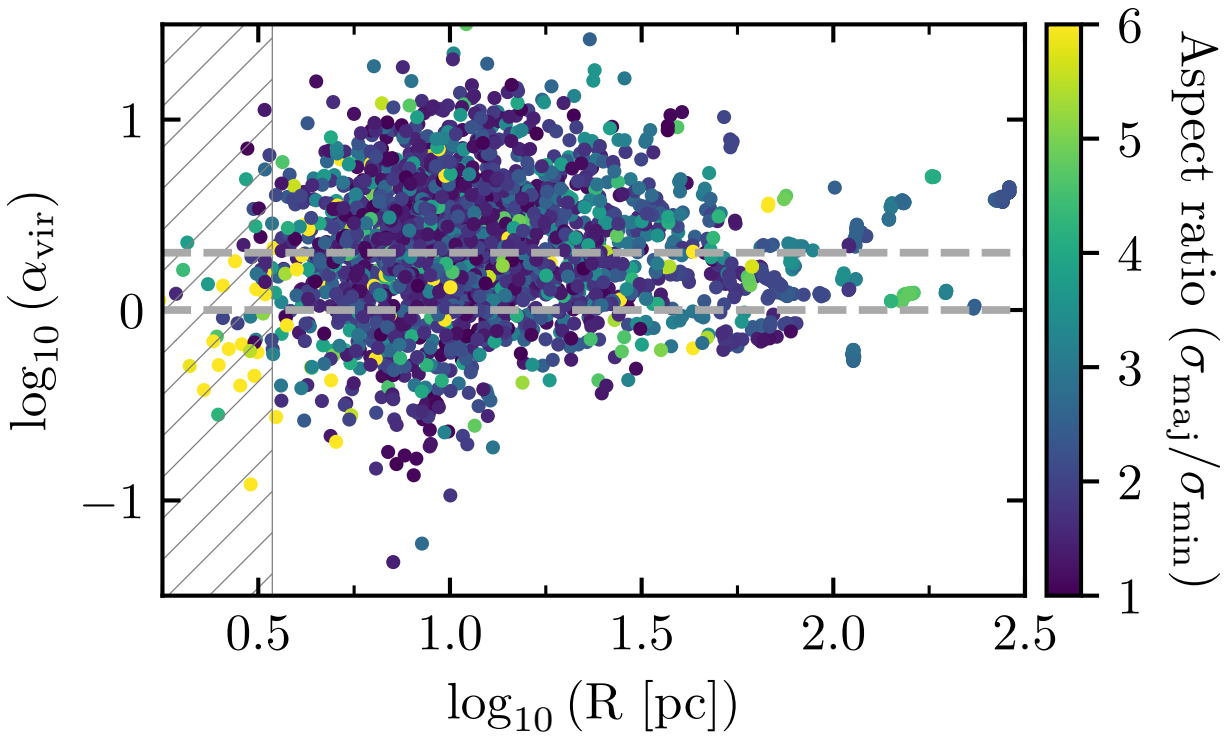

**Figure 15.** Logarithm of the virial parameter against structure radius for all structures in the dendrogram, colored by aspect ratio. The horizontal lines and hatched region are as in Figure 4.

2016, 2022; Y.-H. Xie et al. 2024). The lack of correlation between aspect ratio and size, level, or height suggests that our measurements are not strongly affected by this process. The *orientation* of aspect ratios provides another important test, as preferential alignment along the galaxy's position angle could suggest our structures are large enough to motivate a more detailed treatment of inclination.

However, the orientation distribution of leaf aspect ratios is nearly random across [0, 180]°, with a median and 16%–84% angle of $104^{+52}_{-76}$°. While we do observe a slight excess of structures aligned with the galaxy's axis—30% more leaves have position angles within 10° of the galaxy's orientation than expected from a random distribution (216 versus 167)—this modest enhancement is only marginally significant given our sample of 1500 leaves.

## Appendix E
## Tracing the Hierarchical Ancestry of Leaves

In previous figures, we have treated the populations of leaves and branches as distinct, without taking into account their hierarchical relations. In Figure 16, we reproduce several key plots while highlighting the lineages of six leaves. By "ancestry," we refer to the set of structures defined by following child to parent until the root structure of the tree is reached. This is somewhat analogous to the `levelprops` decomposition discussed in Section 4.4, but treating each hierarchical structure, rather than intensity steps within a single structure, as a level.

The leaves we show were selected to be relatively high in the local hierarchy of their trunks, while representing distinct major branches of the dendrogram (panel (a)). In panels (b)–(g), we present the ancestries of these six leaves traced along the relations of the virial parameter against physical and hierarchical properties as in Figures 4, 5, and 6. We note that several ancestries (e.g., pink, yellow, and purple in Figure 16) belong to the same major trunk, and therefore merge into the same progenitor branch at some point in their hierarchy.

This exercise reveals several interesting conclusions that inform the interpretation of our results:

1. Virial parameters do not strongly correlate with the properties of the originating trunk. In all panels, the properties of individual ancestries appear to decouple from their progenitor trunks after just a few mergers, suggesting that the local properties of dense clumps are not strongly driven by their larger, diffuse envelopes.
2. The majority of leaves (all except yellow) have lower $\alpha_{\rm vir}$ than their ancestors, but these changes do not appear to be monotonic. This is similar to our observations *within* individual structures, discussed in Section 4.4.1.
3. From panel (g), it is evident that the most dramatic changes in the value of $\alpha_{\rm vir}$ (whether up or down) occur near the top of the hierarchy.
4. While significant increases and decreases in $\alpha_{\rm vir}$ are common, the behavior is not consistent across ancestries even for those within the same trunk (e.g., orange and green). Therefore, it is unlikely that our larger analysis, by plotting the sample in its entirety, is obscuring characteristic scales shared within individual trunks.

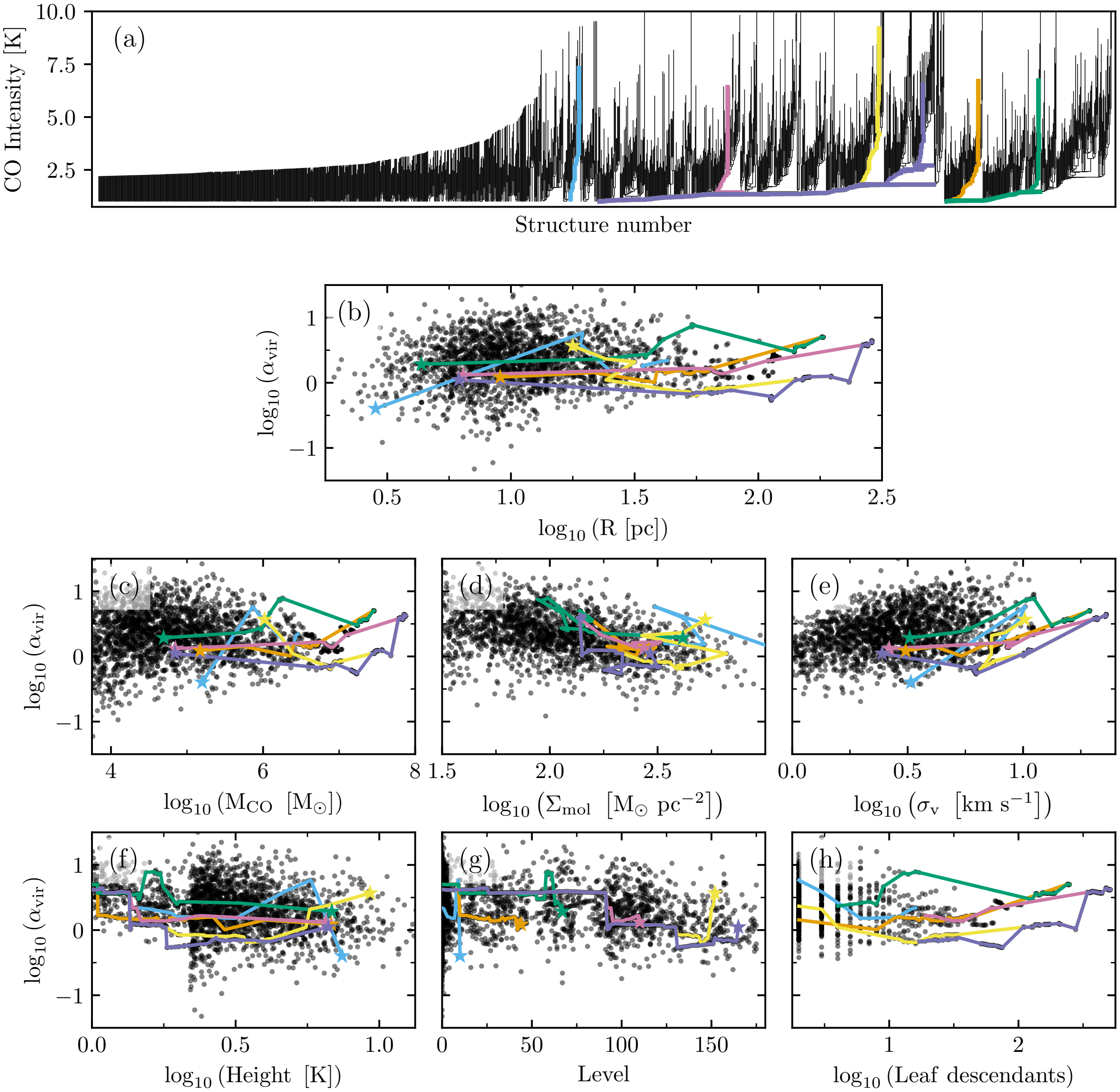

**Figure 16.** Tracing the full ancestry of six leaves through several plots presented in this work. In panels (b)–(h), the six leaves are shown with ⋆ symbols, while their ancestors are traced with a colored line. In some cases, the full hierarchy is obscured by ancestries merging to a common progenitor. (a) Ancestries highlighted on the full dendrogram as in Figure 2. (b) Highlighted on the $\alpha_{\rm vir}-R$ relation as in Figure 4. (c)–(e) Highlighted on the relations of $\alpha_{\rm vir}$ against physical properties (mass, surface density, and velocity dispersion) as in Figure 5. (f)–(h) Highlighted on the relations of $\alpha_{\rm vir}$ against hierarchical properties (height, level, number of leaf descendants) as in Figure 6.

## ORCID iDs

Elias K. Oakes https://orcid.org/0000-0002-0119-1115
Christopher M. Faesi https://orcid.org/0000-0001-5310-467X
Erik Rosolowsky https://orcid.org/0000-0002-5204-2259
Adam K. Leroy https://orcid.org/0000-0002-2545-1700
Simon C. O. Glover https://orcid.org/0000-0001-6708-1317
Annie Hughes https://orcid.org/0000-0002-9181-1161
Sharon E. Meidt https://orcid.org/0000-0002-6118-4048
Eva Schinnerer https://orcid.org/0000-0002-3933-7677
Jiayi Sun https://orcid.org/0000-0003-0378-4667
Amirnezam Amiri https://orcid.org/0000-0002-8553-1964
Ashley T. Barnes https://orcid.org/0000-0003-0410-4504
Zein Bazzi https://orcid.org/0009-0001-1221-0975
Ivana Bešlić https://orcid.org/0000-0003-0583-7363
Frank Bigiel https://orcid.org/0000-0003-0166-9745
Guillermo A. Blanc https://orcid.org/0000-0003-4218-3944
Charlie Burton https://orcid.org/0000-0003-3128-6542
Ryan Chown https://orcid.org/0000-0001-8241-7704
Enrico Congiu https://orcid.org/0000-0002-8549-4083
Daniel A. Dale https://orcid.org/0000-0002-5782-9093
Simthembile Dlamini https://orcid.org/0000-0002-2885-6172
Hao He https://orcid.org/0000-0001-9020-1858
Eric W. Koch https://orcid.org/0000-0001-9605-780X
Fu-Heng Liang https://orcid.org/0000-0003-2496-1247
Jérôme Pety https://orcid.org/0000-0003-3061-6546
Miguel Querejeta https://orcid.org/0000-0002-0472-1011
Sumit K. Sarbadhicary https://orcid.org/0000-0002-4781-7291
Sophia K. Stuber https://orcid.org/0000-0002-9333-387X

Antonio Usero https://orcid.org/0000-0003-1242-505X
Thomas G. Williams https://orcid.org/0000-0002-0012-2142